\documentclass[aps,prd,twocolumn,superscriptaddress]{revtex4-2}

\usepackage[utf8]{inputenc}
\usepackage[T1]{fontenc}
\usepackage{amsmath}
\usepackage{amssymb}
\usepackage{xspace}
\usepackage[margin=2cm]{geometry}
\usepackage[dvipsnames]{xcolor}
\usepackage{graphicx}
\usepackage[sort&compress]{natbib}
\usepackage{changes}
\usepackage{subcaption}
\usepackage{slashed}

\newcommand{\ket}[1]{\ensuremath{|#1\rangle}\xspace}
\newcommand{\bra}[1]{\ensuremath{\langle #1|}\xspace}

\newcounter{comment}

\newcommand{\HMcolor}{red}

\newcommand{\CMcolor}{orange}

            \newcommand{\MDcolor}{BrickRed}

\definecolor{DarkGreen}{RGB}{0,105,62}
\newcommand{\JMMCcolor}{DarkGreen}

\newcommand{\PPcolor}{blue}

\newcommand{\VBcolor}{brown}

\newcommand{\FSBcolor}{violet}

\newcommand{\JScolor}{purple}

{\refstepcounter{comment}%
\begin{quote}
\color{\JMMCcolor}\small$\blacksquare$ \textbf{\underline{Comment} $\sharp$\thecomment.~JMMC:}}%
{\end{quote}}

{\refstepcounter{comment}%
\begin{quote}
\color{\HMcolor}\small$\blacksquare$ \textbf{\underline{Comment} $\sharp$\thecomment.~HM:}}%
{\end{quote}}

{\refstepcounter{comment}%
\begin{quote}
\color{\MDcolor}\small$\blacksquare$ \textbf{\underline{Comment} $\sharp$\thecomment.~MD:}}%
{\end{quote}}

{\refstepcounter{comment}%
\begin{quote}
\color{\CMcolor}\small$\blacksquare$ \textbf{\underline{Comment} $\sharp$\thecomment.~CM:}}%
{\end{quote}}

{\refstepcounter{comment}%
\begin{quote}
\color{\PPcolor}\small$\blacksquare$ \textbf{\underline{Comment} $\sharp$\thecomment.~PP:}}%
{\end{quote}}

{\refstepcounter{comment}%
\begin{quote}
\color{\VBcolor}\small$\blacksquare$ \textbf{\underline{Comment} $\sharp$\thecomment.~VB:}}%
{\end{quote}}

{\refstepcounter{comment}%
\begin{quote}
\color{\JScolor}\small$\blacksquare$ \textbf{\underline{Comment} $\sharp$\thecomment.~JS:}}%
{\end{quote}}

{\refstepcounter{comment}%
\begin{quote}
\color{\FSBcolor}\small$\blacksquare$ \textbf{\underline{Comment} $\sharp$\thecomment.~FSB:}}%
{\end{quote}}

\definechangesauthor[color=\HMcolor]{HM}
\definechangesauthor[color=\CMcolor]{CM}
\definechangesauthor[color=\MDcolor]{MD}
\definechangesauthor[color=\JMMCcolor]{JMMC}
\definechangesauthor[color=\PPcolor]{PP}
\definechangesauthor[color=\VBcolor]{VB}
\definechangesauthor[color=\JScolor]{JS}
\definechangesauthor[color=\FSBcolor]{FSB}

\begin{document}


\title{Pion GPDs: A path toward phenomenology}



\author{José Manuel Morgado Chavez}
\email{josemanuel.morgado@dci.uhu.es}
\affiliation{Department of Integrated Sciences and Center for Advanced Studies in Physics, Mathematics and Computation, University of Huelva, E-21071 Huelva, Spain}

\author{Valerio Bertone}
\email{valerio.bertone@cea.fr}
\affiliation{Irfu, CEA, Université Paris-Saclay, F-91191, Gif-sur-Yvette, France}

\author{Feliciano De Soto Borrero}
\email{fcsotbor@upo.es}
\affiliation{Dpto. Sistemas F\'isicos, Qu\'imicos y Naturales, Universidad Pablo de Olavide, E-41013 Sevilla, Spain}

\author{Maxime Defurne}
\email{maxime.defurne@cea.fr}
\author{Cédric Mezrag}
\email{cedric.mezrag@cea.fr}
\author{Hervé Moutarde}
\email{herve.moutarde@cea.fr}
\affiliation{Irfu, CEA, Université Paris-Saclay, F-91191, Gif-sur-Yvette, France}

\author{José Rodr\'iguez-Quintero}
\email{jose.rodriguez@dfaie.uhu.es}
\affiliation{Department of Integrated Sciences and Center for Advanced Studies in Physics, Mathematics and Computation, University of Huelva, E-21071 Huelva, Spain}

\author{Jorge Segovia}
\email{jsegovia@upo.es}
\affiliation{Dpto. Sistemas F\'isicos, Qu\'imicos y Naturales, Universidad Pablo de Olavide, E-41013 Sevilla, Spain}


\date{\today}

\begin{abstract}
We introduce a new family of Generalised Parton Distribution models able to fulfil by construction all the theoretical properties imposed by QCD. These models are built on standard Parton Distribution Functions and extended to off-forward kinematics through a well-defined procedure. We apply this strategy on the pion, first handling a simple but insightful algebraic model, and then exploiting state-of-the-art computations obtained in continuum QCD. We compare these models with a more standard one relying on an xFitter extraction of the pion Parton Distribution Functions. The results for both quark and gluon Generalised Parton Distributions are presented and exploited for calculation of Electromagnetic, Gravitational and Compton Form Factors. Results on the latter highlight the relevance of next-to-leading order corrections, even in the so-called valence region.
\end{abstract}


\maketitle

\section{Introduction}
\label{sec:Introduction}

Generalised Parton Distributions (GPDs) were introduced more than two decades ago \cite{Mueller:1998fv,Ji:1996nm, Ji:1996ek, Radyushkin:1996ru,Radyushkin:1997ki} and have been deeply studied both theoretically and experimentally since then (see \emph{e.g.} the review papers \cite{Diehl:2003ny,Belitsky:2005qn,Kumericki:2016ehc}). Besides the fact that they allow to describe multiple exclusive processes such as Deeply Virtual Compton Scattering (DVCS) or Deeply Virtual Meson Production (DVMP), the scientific interest for GPDs is fuelled both by the access they offer to the three-dimensional picture of hadrons (tomography) \cite{Burkardt:2000za} and by their connection with the energy-momentum tensor \cite{Ji:1996ek}. The latter allowing in principle an experimental access to the spin decomposition of hadrons on the one hand; and to the pressure and shear forces of partons inside hadrons \cite{Polyakov:2002yz} on the other hand.

As interesting as they are, GPDs are notoriously difficult both to extract from experimental data and to compute using non-perturbative techniques. On the experimental side, exclusive processes related to GPDs are challenging to measure precisely due to the requirement that the hadronic target should not break and therefore require a high luminosity. This has consequences on GPD phenomenology as, for spin $1/2$ targets, the number of independent observables is usually not large enough to properly constrain all amplitudes coming from the different GPDs allowed. This problem is thought to be tamed on spin-0 targets, leading to studies of exclusive processes on $^4\textrm{He}$ targets. Nevertheless, irrespectively of the target's spin, the main channels accessible experimentally do not allow by themselves for an unambiguous extraction of GPDs \cite{Bertone:2021yyz,Bertone:2021wib}. This emphasises the need of non-perturbative QCD computations of GPDs.

On the theoretical side, mainly two ways are used today to compute GPDs: lattice-QCD and Continuum Schwinger Methods (CSM). Following the emergence in the last decade of techniques allowing to extrapolate euclidean results onto the lightfront \cite{Ji:2013dva,Ma:2014jla,Radyushkin:2017cyf,Chambers:2017dov} (see also \cite{Cichy:2018mum,Constantinou:2020hdm}), lattice-QCD practitioners have managed to provide information on the shape of Parton Distribution Functions (PDFs) beyond the computations of their first Mellin moments. Attempts to apply this techniques to GPDs are still ongoing, the first results at vanishing skewness on the pion being encouraging \cite{Chen:2019lcm}. However, going beyond vanishing skewness remains extremely challenging due to the set of theoretical constraints that GPDs need to obey \cite{Alexandrou:2020zbe}. Nevertheless, lattice-QCD is certainly a promising path to be followed in the future.

The last decade was also very fruitful for CSM practitioners, on the path to compute GPDs. After early attempts using diagrammatic representations \cite{Chang:2014lva,Mezrag:2014tva,Mezrag:2014jka,Mezrag:2016hnp}, the path through Light-Front Wave Functions (LFWFs) opened the possibility to obtain GPDs fulfilling all the required theoretical properties \cite{Chouika:2017dhe,Chouika:2017rzs}. However, if a proof of principle has been achieved, the most refined continuum techniques remain to be used in the case of GPDs. Nonetheless, they have very recently been used to evaluate consistently both the pion's Parton Distribution Amplitude (PDA) and PDFs \cite{Ding:2019lwe,Cui:2020tdf} (see also \cite{Chang:2021utv} for a review of recent results on the pion) and first steps toward GPDs have been undertaken \cite{Shi:2020pqe,Raya:2021zrz,Zhang:2021mtn}.

Among hadrons, the pion has been one of the main topics of CSM studies for several reasons: i) it presents the characteristic of being both a QCD bound-state and a Goldstone boson of chiral symmetry. ii) Because of this double role, a description consistent with experimental data requires a proper treatment of symmetries \cite{Binosi:2016rxz}, making it an ideal test-ground for new techniques; and iii), presenting a two-body leading Fock state, it remains simpler in some aspects than the nucleon. These CSM studies have contributed to push forward experimental studies such as the extraction of Electromagnetic Form Factors (EFFs) at large momentum transfer or the pion PDF through tagged Deep Inelastic Scattering (tDIS) (see \emph{e.g} \cite{Aguilar:2019teb}). 

In view of this renewed experimental interest, especially in the perspective of future electron-ion colliders in the USA (EIC) and in China (EicC), we present in this paper the first CSM-based pion-GPD model able to fulfil by construction all the required theoretical constraints. In order to make it relevant for phenomenological purposes, we take advantage of state-of-the-art pion PDF computations based on CSM, together with the latest experimental extractions of EFFs \cite{Huber:2008id} and Gravitational Form Factors (GFFs) \cite{Kumano:2017lhr}. We compare this model with more standard approaches based on the Radyushkin Double Distribution Ansatz (RDDA) \cite{Musatov:1999xp}. Thus we start in section II by recalling the reader the set of theoretical constraints that GPDs have to obey and the consequences of the latter. In section III we present our modelling strategy in a general fashion, and exploit it to develop two pion's GPD models within the DGLAP region, one of which is based on the PDF of Ref. \cite{Ding:2019lwe}. In section IV we present our way to extend the model to the so-called ERBL region, improving the numerical techniques introduced in \cite{Chouika:2017dhe}. Section V is devoted to the development of a RDDA-based phenomenological model. Finally, in section VI, we compute and discuss DVCS Compton Form Factors (CFFs) at next-to-leading order (NLO) using the three models presented.

\section{Definition and properties of GPDs}
\label{sec:Definition and Properties}

In this section we remind the reader the definition of the pion GPDs and present all the properties they should fulfil together with their consequences.

\subsection{Definition and properties}
\label{subsec:Definition and Properties}

GPDs are defined from the Fourier transform of non-local hadronic matrix elements. Their number depends on the spin of the considered hadron. For the pion, which will be our main focus throughout this paper, one has \cite{Diehl:2003ny}:
\begin{align}
\label{eq:quarkGPDDef}
& H^{q}_{\pi}\left(x,\xi,t\right) = \hspace{-0.1cm}\int\frac{d\lambda}{2\pi}e^{ix\lambda}\bra{p'}\bar{\psi}^{q}\left(\hspace{-0.1cm}-\frac{\lambda n}{2}\right)\slashed{n}\psi^{q}\left(\frac{\lambda n}{2}\right)\ket{p}\\
\label{eq:gluonGPDDef}
& H^{g}_{\pi}\left(x,\xi,t\right) = \hspace{-0.1cm}\displaystyle \int\frac{d\lambda}{2\pi}e^{ix\lambda}\bra{p'}G^{\mu}_{\alpha}\left(\hspace{-0.1cm}-\frac{\lambda n}{2}\right)G^{\beta}_{\mu}\left(\frac{\lambda n}{2}\right)\ket{p}n^{\alpha}n_{\beta}
\end{align}
where, $\psi^q$ is a quark field of a given flavour $q$; $G^{\mu\nu}$ is the gluon field strength and ``$n$'' is a light-like four-vector normalised such that $n\cdot P=1$ with $P = (p+p')/2$. Note that for brevity we have omitted the dependence on the renormalisation scale $\mu$, controlled by evolution equations, and the expression of the Wilson line. The average momentum fraction carried by the active parton is labelled $x$. The skewness $\xi =\left[\left(p-p'\right)\cdot n \right]/\left[\left(p'+p\right) \cdot n\right]$ corresponds to the momentum fraction exchanged along the lightcone, and $t$ stands for the standard Mandelstam variable. 

GPDs are defined for $x\in\left[-1,1\right]$ \cite{Diehl:1998sm} and $\xi\in\left[-1,1\right]$. Continuation to $\xi\in\left[1,\infty\right)$ is possible through Generalised Distribution Amplitudes thanks to crossing symmetry \cite{Diehl:1998dk,Diehl:2000uv}. Time reversal invariance and hermiticity guarantee that pion GPDs are real functions and even in $\xi$ \cite{Diehl:2003ny,Belitsky:2005qn}.

In the forward limit, \emph{i.e.} when both $\xi$ and $t$ vanish, GPDs reduce to partons distribution functions (PDFs):
\begin{align}
  \label{eq:ForwardLimitQuark}
  &H^{q}\left(x,0,0\right) = q\left(x\right)\Theta\left(x\right) -\bar{q}\left(-x\right)\Theta\left(-x\right), \\
  \label{eq:ForwardLimitGluon}
  &H^{g}\left(x,0,0\right) = x g\left(x\right)\Theta\left(x\right) -xg\left(-x\right)\Theta\left(-x\right),
\end{align}
where $\Theta$ is the Heaviside distribution, $q(x)$ is the quark PDF of flavour $q$, $\bar{q}(x)$ is the anti-quark PDF, and $g(x)$ the gluon PDF. 

Beyond the forward limit, PDFs also constrain the GPDs at non-vanishing $\xi$ through the so-called positivity bounds. The latter come from the underlying Hilbert-space structure and, in the case of the pion, state that \cite{Pire:1998nw,Radyushkin:1998es,Diehl:2000xz,Pobylitsa:2002gw}:
\begin{align}
  \label{eq:PositivityQuark}
  &\left|H^{q}_{\pi}\left(x,\xi,t\right)\right| \leq \sqrt{q\left(x_{in}\right)q\left(x_{out}\right)}\\
  \label{eq:PositivityGluon}
  &\left|H^{g}_{\pi}\left(x,\xi,t\right)\right| \leq \sqrt{(1-\xi^2)}\sqrt{x_{in}x_{out}g\left(x_{in}\right)g\left(x_{out}\right)}
\end{align}
with
\begin{align}
  \label{eq:xinxout}
  x_{in}= \frac{x+\xi}{1+\xi}, \quad x_{out} = \frac{x-\xi}{1-\xi},
\end{align}
in the so-called DGLAP (or outer) kinematic region $(|x|\ge |\xi|)$.
Such a property has been shown to be stable under leading logarithm evolution \cite{Pobylitsa:2002iu}.

Because of the historical difficulty in fulfilling both at the same time, the positivity property is usually put in parallel to another important property called polynomiality. It follows as a direct consequence of Lorentz covariance and states that the Mellin moments of GPDs are polynomials in $\xi$ \cite{Ji:1998pc,Radyushkin:1998bz,Polyakov:1999gs}:
\begin{align}
  \label{eq:PolynomialityQuark}
  \int_{-1}^1 \hspace{-0.1cm} dx x^n H^q(x,\xi) &= \hspace{-0.1cm}\sum_{i=0}^{[n/2]}(2\xi)^{2i}A^q_{n+1,2i} \nonumber \\
  & \quad \quad + \textrm{mod}(2,n)(2\xi)^{n+1}C^q_{n+1}\\
  \label{eq:PolynomialityGluon}
  \int_{0}^1 \hspace{-0.1cm} dx x^{n-1} H^g(x,\xi) &= \hspace{-0.1cm}\sum_{i=0}^{[n/2]}(2\xi)^{2i}A^g_{n+1,2i}+(2\xi)^{n+1}C^g_{n+1}
\end{align}
where, in the gluon case, $n$ is always odd as the gluon-GPD is even. Here, $\left[\cdot\right]$ represents the ``floor function''. We note also that in the case $n=0$, the Mellin moment of the quark GPD yields the pion's EFF.

Even if GPDs are matrix elements encoding non-perturbative information on hadron's structure, perturbative QCD (pQCD) still provides us with constraints on GPDs. We highlight here the pion's GPD behaviour when $x\to 1$ \cite{Yuan:2003fs}:
\begin{align}
  \label{eq:LargeX}
  H^{q}_{\pi}\left(x,\xi,t\right) \sim \frac{\left(1-x\right)^2}{1-\xi^2}
\end{align}

Furthermore, when $-t$ becomes large, GPDs can be expressed in terms of a convolution of distribution amplitudes with perturbative kernels \cite{Hoodbhoy:2003uu}, generalising the seminal results of the electromagnetic form factor \cite{Lepage:1980fj,Efremov:1979qk}. Up to logarithmic corrections, one obtains:
\begin{align}
  \label{eq:larget}
  H^q\left(x,\xi,t\right) = \frac{1}{-t}f^q\left(x,\xi,\alpha_{\text{S}}\left(t\right)\right)
\end{align}
where $\alpha_s$ is the strong coupling constant and $f^{q}\left(x,\xi,\alpha_{\text{S}}\left(t\right)\right)$ simply labels the limit $\substack{\lim\\ -t\to\infty} -tH\left(x,\xi,t\right)$ (see \emph{e.g.} \cite{Melic:1998qr} for an extensive discussion in the case of the pion form factor).

Finally, pQCD also constrains mathematical properties of GPDs through their connections with experimental processes. Sometimes called the ``Golden GPD channel'', the amplitude of DVCS can be factorised in terms of coefficient functions, calculable in perturbation theory, and GPDs \cite{Collins:1998be,Radyushkin:1997ki,Ji:1998xh}. Yet, the factorised amplitude is finite only if the GPDs are continuous on the line $x=\pm \xi$ (this is also true for other processes). At this point, considering the GPDs as a hadron-parton scattering amplitude, Collins and Freund \cite{Collins:1998be} showed that the latter needs to be continuous, but non-analytic, on the lines $x=\pm\xi$. This is compatible with the one-loop evolution equations (see \emph{e.g.} \cite{EvolutionPaper}).

Last but not least, since our study is focused on pion GPDs, we must mention the so-called soft-pion theorem \cite{Polyakov:1998ze,Mezrag:2014jka}. This property tells us that:
\begin{align}
  \label{eq:SoftPionTheorem}
  H^{q}_{\pi}\left(x,1,0\right) = \frac{1}{2}\varphi_{\pi}\left(\frac{1+x}{2}\right),
\end{align}
where $\varphi_{\pi}$ is the leading-twist pion PDA.

The list above shed light on the properties obeyed by GPDs. One should note that before the present paper, no realistic model was able to fulfil all these constraints by construction. The first lattice computation at non-vanishing $\xi$ \cite{Alexandrou:2020zbe} also fails to do so as the support, continuity and large-$x$ behaviour properties are violated in such a pioneering computation.

\subsection{Double distributions}
\label{sec:DD}

Among all the GPDs theoretical properties enumerated above, polynomiality holds a special place. Indeed, introducing an odd function $D(z)$ for $z \in [-1,1]$, called the D-term, such that:
\begin{align}
  \label{eq:DtermGeneration}
  \int_{-1}^1 dz z^n D\left(z\right) = \textrm{mod}(2,n) 2^n C_{n+1},
\end{align}
one can show \cite{Chouika:2017dhe} that, for each $t$, $H\left(x,\xi\right)-\textrm{sign} (\xi) D\left(x/\xi\right)$ fulfils the so-called Lugwig-Helgason consistency condition \cite{Hertle:1983} (called Cavalieri condition in \cite{Teryaev:2001qm}) meaning that $H-D$ is in the range of the Radon transform. More precisely:
\begin{align}
  \label{eq:RadonTransformDef}
  H\left(x,\xi\right) &= \textrm{sign}\left(\xi\right) D\left(\frac{x}{\xi}\right) + \int d\Omega F\left(\beta,\alpha\right) \delta\left(x-\beta-\xi\alpha\right) \nonumber \\
  &= \int d\Omega \left[F\left(\beta,\alpha\right)+\xi \delta\left(\beta\right)D\left(\alpha\right)\right] \delta\left(x-\beta-\xi\alpha\right)
\end{align}
with $d\Omega = d\beta d\alpha \Theta\left(1-\left|\beta\right|-\left|\alpha\right|\right)$. The reader may recognise the Double Distribution (DD) introduced by Radyushkin \cite{Radyushkin:1997ki} (also called spectral functions in \cite{Mueller:1998fv}):
\begin{align}
  \label{eq:DDDef}
  H\left(x,\xi\right) = \int d\Omega \left[ F\left(\beta,\alpha\right) + \xi G\left(\beta,\alpha\right)\right]\delta\left(x-\beta-\alpha \xi\right),
\end{align}
in the so-called Polyakov-Weiss scheme \cite{Polyakov:1999gs} where $G$ is reduced to the D-term times a Dirac delta.

Relation \eqref{eq:DDDef} between DDs and GPDs is scheme-independent (see \cite{Chouika:2017dhe} for a detailed description). For instance, in the P-scheme \cite{Pobylitsa:2002vi} (which will be useful in the next sections), $\left(F,G\right)$ DDs are redefined as: 
\begin{align}
    \label{eq:PschemeF}
    F(\beta,\alpha) &= (1-|\beta|)h_p(\beta,\alpha), \\
    \label{eq:PschemeG}
    G(\beta,\alpha) &= -\textrm{sign}(\beta) \alpha h_p (\beta,\alpha),
  \end{align}
where $h_p$ is often called a DD by abuse of terminology. The reader interested in the formulae allowing to go from one scheme to another is referred to Ref. \cite{Chouika:2017dhe} and references therein

We highlight that, from DDs in any scheme, one can recover two scheme-independent quantities, the PDF $q(x)$ and the so-called D-term through:
\begin{align}
  \label{eq:DDtoPDF}
  q(\beta) = \int_{-1+|\beta|}^{1-|\beta|} \textrm{d}\alpha F(\beta,\alpha)\, , \\
  \label{eq:DDtoD}
  D(\alpha) = \int_{-1+|\alpha|}^{1-|\alpha|} \textrm{d}\beta G(\beta,\alpha)\, .
\end{align}

The theoretical constraints fulfilled by GPDs find their analogues when handling DDs. The parity in $\xi$ becomes parity in $\alpha$; the DDs support guarantees the GPDs one, and the continuity  on the $|x|=|\xi|$ lines is embedded in the behaviour of the DDs at the points $(\beta,\alpha)=(0,\pm1)$. Apart from a seminal paper on the topic \cite{Pobylitsa:2002vi}, little work has been done on expressing the positivity property in the DDs space. Nevertheless, this does not preclude exploiting DDs in order to build GPD models fulfilling by construction all the required theoretical properties as we will see below.

\section{Modelling the DGLAP region of the pion GPDs}
\label{sec:DGLAPRegion}

Following previous results showing how to exploit the Radon transform relation between GPDs and DDs, we adopt the so-called covariant extension strategy \cite{Chouika:2017dhe,Chouika:2017rzs} to develop a brand new program for modelling pion GPDs fulfilling all of the fundamental properties required by the underlying quantum field theory (see Sec. \ref{subsec:Definition and Properties}). It consists in modelling the GPDs within the DGLAP region in such a way that the positivity property is fulfilled. Then, the use of the inverse Radon transform allows us to obtain the associated DD and consequently build the inner (or ERBL) region ($\left|x\right|\leq\left|\xi\right|$) through the Radon transform (see Eq. \eqref{eq:DDDef}). The polynomiality property of the resulting GPD is therefore guaranteed by construction.

In this section we focus on the first step of this procedure, presenting a general approach for modelling of DGLAP quark-GPDs with a built-in positive character. We exploit it to develop a whole new family of DGLAP GPDs and illustrate it with an alternative derivation of the algebraic model presented in \cite{Chouika:2017rzs}, which we extensively use as a benchmark. Finally, we introduce a brand-new GPD model based on forefront CSM studies.

\subsection{General framework}

Among the possible ways of fulfilling the positivity property when modelling GPDs, we choose to exploit the overlap of light-front wave functions \cite{Diehl:2003ny, Diehl:2000xz}, mainly for two reasons: i) it provides a desirable probabilistic interpretation, similarly to non-relativistic quantum mechanics \cite{Brodsky:1997de,Brodsky:1997de, Diehl:2000xz, Diehl:2002he, Brodsky:2000xy}; and ii), non-perturbative computations of these LFWFs have been performed using different techniques providing, at least in principle, a connection to QCD. However, a difficulty arises; namely that the type of overlap varies with the kinematical regions. Only the so-called DGLAP region can be described through an overlap of LFWFs with the same number of partons. There, the overlap has the structure of a scalar product in an Hilbert space, guaranteeing by construction the fulfilment of the positivity property \cite{Diehl:2000xz}.



LFWFs depend on two types of kinematic variables: $x_i$, representing the longitudinal momentum fraction of the hadron's average lightcone momentum carried by a given parton; and $\vec{k}^{\perp}_i$, its momentum in the transverse-plane (defined with respect to the hadron's momentum in the infinite momentum frame). In addition, they depend on a renormalisation scale, $\mu$.

Longitudinal momentum fractions for each active parton are defined in Eq. \eqref{eq:xinxout}. Transverse momentum fractions are defined analogously:
\begin{equation}\label{eq:kout}
\begin{array}{rcl}
	\displaystyle \vec{k}^{\perp}_{out} & \displaystyle = & \displaystyle \vec{k}_{\perp}+\left(1-x_{out}\right)\frac{\vec{\Delta}_{\perp}}{2}\quad\text{,}
\end{array}
\end{equation}\vspace{-0.5cm}
\begin{equation}\label{eq:kin}
\begin{array}{rcl}
	\displaystyle \vec{k}^{\perp}_{in} & \displaystyle = & \displaystyle \vec{k}_{\perp}-\left(1-x_{in}\right)\frac{\vec{\Delta}_{\perp}}{2}
\end{array}
\end{equation}

\noindent
with $\vec{\Delta}_{\perp}$ the momentum transfer in the hadron's transverse plane.

A complete description of GPDs using LFWFs requires the knowledge of an infinite set of the latter. Thus, the path to GPD modelling from LFWFs requires assumptions. The one we choose is to truncate the pion's state Fock-space expansion to its leading component. We assume that, at a low enough energy scale, a description of hadrons in terms of valence dressed-quarks (not partons) is a reasonable approximation; which could, in principle, be systematically improved using LFWFs of higher number of dressed-quarks and gluons. Under such an assumption, the infinite sum involved in the overlap representation of GPDs for a meson can be truncated to its first term, writing:
\begin{align}\label{eq:GPDOverlapRepTruncated}
 &\left.H^{q}_{h}\left(x,\xi,t;\mu_{\text{Ref.}}\right)\right|_{\left|x\right|\geq\left|\xi\right|}= \nonumber \\
 &\int\frac{d^{2}k_{\perp}}{16\pi^{3}}\Psi^{*}_{q/h}\left(x_{out},\vec{k}^{\perp}_{out};\mu_{\text{Ref.}}\right)\Psi_{q/h}\left(x_{in},\vec{k}^{\perp}_{in};\mu_{\text{Ref.}}\right)
\end{align}
with $\Psi$ labelling the two-body LFWFs, and where a sum over flavour, helicity and colour sates is implicitly understood. Such a truncation preserves the structure of the Hilbert space, conserving therefore the positivity property.

The reader may argue that the computation of such valence LFWFs for mesons is already a hard task by itself. There are different ways to proceed; maybe, the easiest course to obtain a description in terms of effective particles being the covariant treatment of the corresponding quantum field theoretical bound-state equations \cite{Eichmann:2016yit, Binosi:2016rxz, Fischer:2018sdj, Qin:2020jig, Mezrag:2020iuo} and its projection onto the light-front.

In this respect, recent studies \cite{Xu:2018eii, Zhang:2021mtn, Raya:2021zrz} suggest that the use of factorised \textit{Ansätze} for LFWFs provides a fair approximation to the description of non-perturbative hadronic features that may be difficult to grasp from first-principle calculations. In particular, the factorisation hypothesis has proved to yield a good approximation in the description of pions \cite{Xu:2018eii}. Thus, as far as we are concerned with a phenomenological study of pions, considering such an approach deserves special attention. 

Therefore, we consider the leading-twist two-particles, \emph{e.g.} $\pi$, LFWF for a pair of quarks of flavour $(q_1,q_2)$ and of helicity $(\lambda_1,\lambda_2)$:
\begin{align}\label{eq:SeparableLFWF}
\Psi^{\lambda_1\lambda_2}_{q_1q_2/\pi}\left(x,\vec{k}_{\perp}\right)=f_{q_1q_2/\pi}\left(x\right)g^{\lambda_1\lambda_2}_{q_1q_2/\pi}\left(k^{2}_{\perp}\right)
\end{align}
where a sum of colour degrees of freedom is understood. For simplicity in the notation, we have omitted any explicit reference to the renormalisation scale $\mu_{\textrm{Ref.}}$.

Our truncation requires the LFWF in \eqref{eq:SeparableLFWF} to satisfy the sum rule,
\begin{align}
  \label{eq:LFWFNormalization}
  q_{\pi}\left(x\right)=\sum_{\substack{
  \lambda_1, \lambda_2 \\
  q_1, q_2}
  }\delta_{qq_1}\int\frac{d^{2}k_{\perp}}{16\pi^{3}}\left|\Psi^{\lambda_1\lambda_2}_{q_1q_2/\pi}\left(x,\vec{k}_{\perp}\right)\right|^{2}
\end{align}
with $q_{\pi}\left(x\right)$ the leading-twist quark PDF.

One can build a simple \emph{Ansatz} for $f_{q_1q_2/\pi}$ in order to fulfil Eq. \eqref{eq:LFWFNormalization}:
\begin{align}
  \label{eq:NormalizedLFWF}
f_{q_1q_2/\pi}(x) = \sqrt{q_{\pi}\left(x\right)}
\end{align}
%
and therefore, employing the formalism of the overlap representation, build two-particle bound-states DGLAP GPDs. Plugging Eq. \eqref{eq:NormalizedLFWF} into \eqref{eq:GPDOverlapRepTruncated}, one gets \cite{Zhang:2021mtn}:
\begin{align}\label{eq:ModellingMasterRelation}
\left.H^{q}_{\pi}\left(x,\xi,t\right)\right|_{x\geq\left|\xi\right|}=\sqrt{q_{\pi}\left(x_{in}\right)q_{\pi}\left(x_{out}\right)}\Phi^{q}_{\pi}\left(x,\xi,t\right)
\end{align}
which defines the ``master equation'' for our modelling strategy. 

The above relation defines a whole new family of DGLAP GPDs, built on a basis of parametrisations for the corresponding hadron PDFs. Furthermore, its momentum-transfer dependence is encoded into a single function,
\begin{align}
  \label{eq:GPDtDependence}
&\Phi^{q}_{\pi}\left(x,\xi,t\right)\nonumber \\
&= \sum_{\substack{
  \lambda_1, \lambda_2 \\
  q_1, q_2}
  }\delta_{qq_{i}}\int\frac{d^{2}k_{\perp}}{16\pi^{3}}g^{\lambda_{1}\lambda_{2}^{*}}_{q_{1}q_{2}/\pi}\left(k^{\perp^{2}}_{in}\right)g^{\lambda_{1}\lambda_{2}}_{q_{1}q_{2}/\pi}\left(k^{\perp^{2}}_{out}\right)
\end{align}
which can be directly obtained from separable \textit{Ansätze} for LFWFs. As a final remark notice that, from such expression one might expect the variables $x$, $\xi$ and $t$ to be correlated; as indicated by both theoretical considerations \cite{Burkardt:2004bv} and Lattice studies \cite{Brommel:2005ee}. This will turn out to be one of the main features exhibited the GPD models presented here in (Eqs. \ref{eq:PhiT}-\ref{eq:zetaDef}).

Interestingly, the canonical normalisation \eqref{eq:LFWFNormalization} of the LFWF \eqref{eq:SeparableLFWF} requires that $\Phi^{q}_{\pi}\left(x,\xi,t=0\right)=1$, and therefore one trivially gets:
\begin{align}\label{eq:SaturationPos}
H^{q}_{\pi}\left(x,\xi,0\right)=\sqrt{q_{\pi}\left(x_{in}\right)q_{\pi}\left(x_{out}\right)}
\end{align}
revealing an interesting property of the present modelling strategy: GPDs built under the assumption of LFWF factorisation \emph{saturate} the positivity property at vanishing $t$.

Finally, let us highlight that the present model provides a flexible and valuable new way to model GPDs at low scale through PDFs, which are usually well known quantities (although not much in the case of the pion). In addition, if the $t$-dependence is ``factorised out'' of the PDF's dependence, it remains intertwined with the momentum fraction $x$ and $\xi$. This is again an interesting alternative to the models presenting a fully factorised $t$-dependence.

\subsection{Pion GPDs}
\label{subsec:PionGPDs}

As mentioned above, obtaining the valence LFWFs is a research topic by itself. Different approaches can be envisioned, like light-front basis techniques or AdS/QCD \cite{deTeramond:2018ecg,Rinaldi:2017roc,Lan:2019rba}. In this work, we will focus on the CSM one \cite{Salpeter:1951sz,Gell-Mann:1951ooy}, briefly describing how the LFWFs can be obtained from the Bethe-Salpeter Wave Function (BSWF). The latter is the solution of the so-called Bethe-Salpeter equation \cite{Salpeter:1951sz,Gell-Mann:1951ooy}, describing bound states in a covariant way. Interestingly, one can project the two-body BSWF of mesons on the lightfront to recover the valence LFWFs.

Light-front projection is unfortunately not that simple, mainly because, using CSM, the solutions are obtained using an euclidean metric, meaning that the lightcone is reached only in continuing the momentum dependence to the complex plane. This difficulty can be bypassed using the Nakanishi representation \cite{nakabook,Nakanishi:1963zz,Nakanishi:1969ph} as exemplified in  Ref. \cite{Chouika:2017rzs}, following earlier studies \cite{Chang:2013pq,Chang:2014lva,Mezrag:2014tva,Mezrag:2014jka,Mezrag:2016hnp,Chen:2016sno,Salme:2017oge,Carbonell:2017kqa,Carbonell:2017isq}. This allowed calculations of the pion's quark-LFWFs with and without orbital angular momentum, from which we take advantage here.

Indeed, exploiting the results of Ref. \cite{Chouika:2017rzs}, and thus taking into account the two independent helicity states of the quarks ($\uparrow \downarrow$ and $\uparrow \uparrow$), we can derive the following expression for the $\Phi$ function of  Eqs. \eqref{eq:ModellingMasterRelation}-\eqref{eq:GPDtDependence}:
\begin{align}
  \label{eq:PhiT}
  \Phi^{q}_{\pi}\left(x,\xi,t\right)=\frac{1}{4}\frac{1}{1+\zeta^2}\left(3+ \frac{1-2\zeta}{1+\zeta}\frac{\textrm{arctanh}\left(\sqrt{\frac{\zeta}{1+\zeta}} \right)}{\sqrt{\frac{\zeta}{1+\zeta}}} \right)
\end{align}
with 
\begin{align}\label{eq:zetaDef}
\zeta = -\frac{t}{4M^2}\frac{(1-x)^2}{1-\xi^2}.
\end{align}

Exploring the effect of various models for the pion PDFs may open the door to a large number of pion GPD models built on the basis of simple fundamental assumptions. In this work we exploit this possibility with two existing parametrisations: a simple one, which has already been reported in \cite{Chouika:2017rzs} for the study of GPDs; and one further, more sophisticated PDF parametrisation based on state-of-the-art CSM approaches to QCD \cite{Ding:2019lwe}.
\begin{figure}[t]
  \centering
  \includegraphics[scale=0.45]{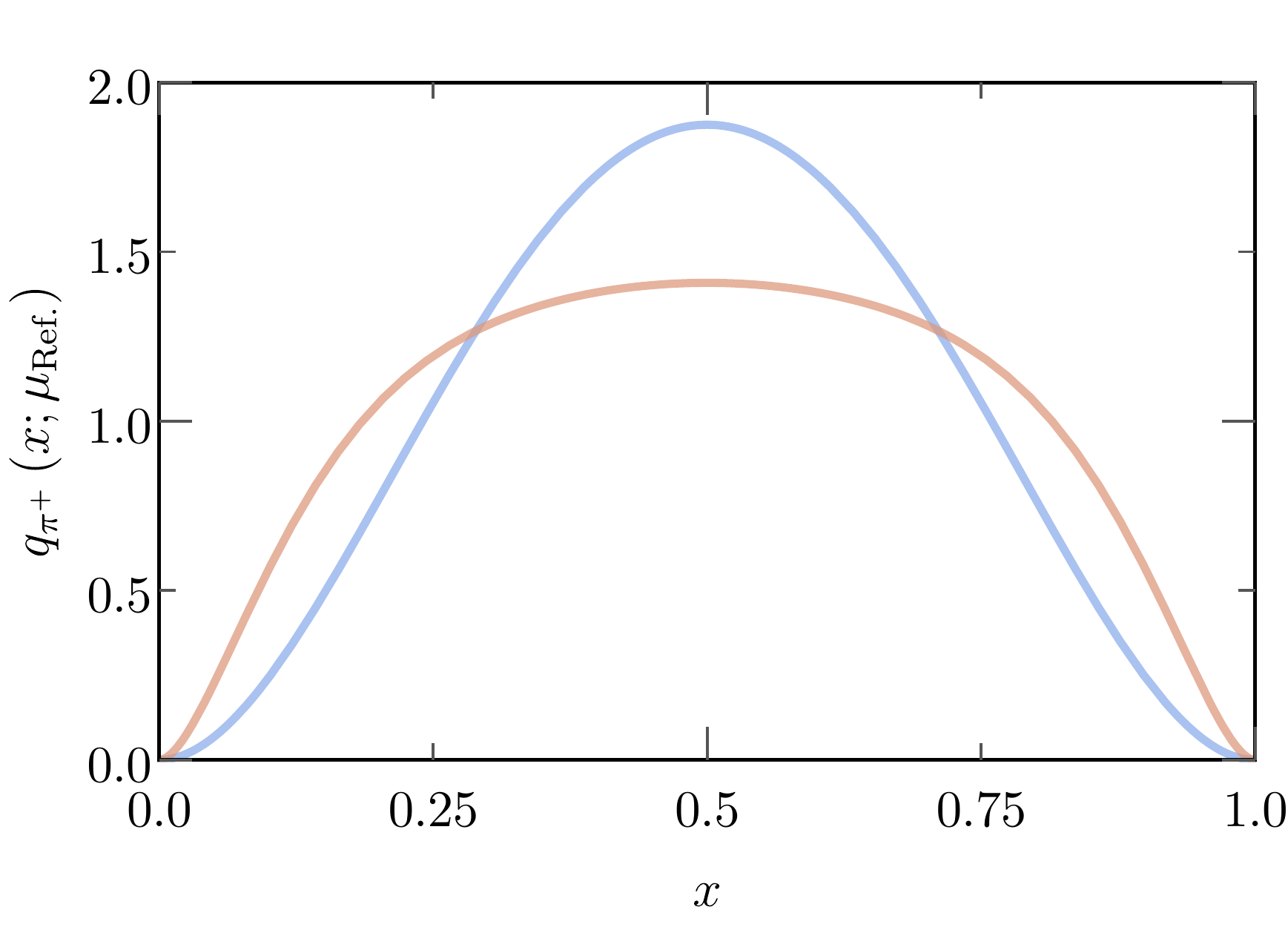}
  \caption{Comparison between the algebraic $u$-quark PDF computed in Ref. \cite{Chouika:2017rzs} (blue line), and the one obtained through the solution of the Bethe-Salpeter equation in Ref. \cite{Cui:2020tdf} (brown line).}
  \label{fig:PdfComparison}
\end{figure}

\subsubsection{Algebraic PDF parametrisation}

A first possible parametrisation for a quark PDF in a pion was given in \cite{Chouika:2017rzs} through (see also Fig. \ref{fig:PdfComparison}):
\begin{equation}\label{eq:AlgebraicModelPDF}
q_{\pi}\left(x\right)=30x^{2}\left(1-x\right)^{2}
\end{equation}
Plugged into Eq. \eqref{eq:ModellingMasterRelation}, such a PDF straightforwardly yields the GPD (Fig. \ref{fig:gpdComparison}):
\begin{align}\label{eq:AlgebraicModelDGLAP}
\left.H^{q}_{\pi}\left(x,\xi,t\right)\right|_{x\geq\left|\xi\right|}^{\text{Alg.}} = 30\frac{\left(1-x\right)^{2}\left(x^{2}-\xi^{2}\right)}{\left(1-\xi^{2}\right)^{2}}\Phi^{q}_{\pi}\left(x,\xi,t\right)
\end{align}
with $\Phi^{q}_{\pi}\left(x,\xi,t\right)$ given in Eq. \eqref{eq:PhiT} and where the mass scale $M$ is fitted to the experimental value of the pion charge radius at a value of $M = \left(318 \pm 4\right) \textrm{MeV}$ (see \cite{Chouika:2017rzs}).

This algebraic parametrisation presents the expected large-$x$ behaviour, but not the asymptotic $1/t$ decrease. Instead a $1/t^2$ decrease is obtained, due to not taking into account all the four components of the BSA (see for instance \cite{Zhang:2021mtn, Raya:2021zrz} for an example on how to correct it). Fortunately, since experimental interest on GPDs holds at low values of $|t|$, the algebraic model results should remain relevant to obtain reasonable experimental yields. And in fact, it is able to reproduce well the available data on pion's electromagnetic form factors, even for $\left|t\right|$ above $1~\textrm{GeV}^2$ \cite{Chouika:2017rzs}.

\subsubsection{Numerical PDF parametrisation}

The quark PDF parametrisation giving rise to the GPD model of Eq. \eqref{eq:AlgebraicModelDGLAP} is known to fail in the description of dynamical chiral symmetry breaking in QCD \cite{Chang:2013pq}. The natural next step is then to employ \textit{Ansäzte} accounting for such a fundamental phenomena of QCD.

In particular we decide to employ the realistic pion PDF presented in \cite{Ding:2019lwe}. There, the authors employed a symmetry-preserving truncation scheme for the system of
Dyson-Schwinger equations that led them to a numerical solution to the Bethe-Salpeter equation which, for the case of quark PDFs, yielded:
\begin{align}\label{eq:NumericalPDF}
q_{\pi}\left(x\right)=\mathcal{N}_{q}x^{2}\left(1-x\right)^{2}\left[1+\gamma x\left(1-x\right)+\rho\sqrt{x\left(1-x\right)}\right]
\end{align}
with $\mathcal{N}_{q}=213.32$, $\gamma=2.2911$ and $\rho=-2.9342$; defined at a reference scale $\mu_{\text{Ref.}}=311\,\text{MeV}$ \cite{Cui:2020tdf}. Notice that such a PDF, as well as that of Eq. \eqref{eq:AlgebraicModelPDF} exhibit the $x\rightarrow 1$ behaviour predicted by QCD's parton model. Also, because of the two-body approximation, the same behaviour appears in the $x\rightarrow 0$ limit.

Fig. \ref{fig:PdfComparison} shows a comparison with the simple PDF of Eq. \eqref{eq:AlgebraicModelPDF}. That figure reveals a crucial difference between both models: owing to dynamical chiral symmetry breaking, the present parametrisation exhibits a shape broader than the algebraic PDF employed through previous subsection \cite{Ding:2019lwe}.

\begin{figure}[t]
  \centering
  \includegraphics[scale=0.45]{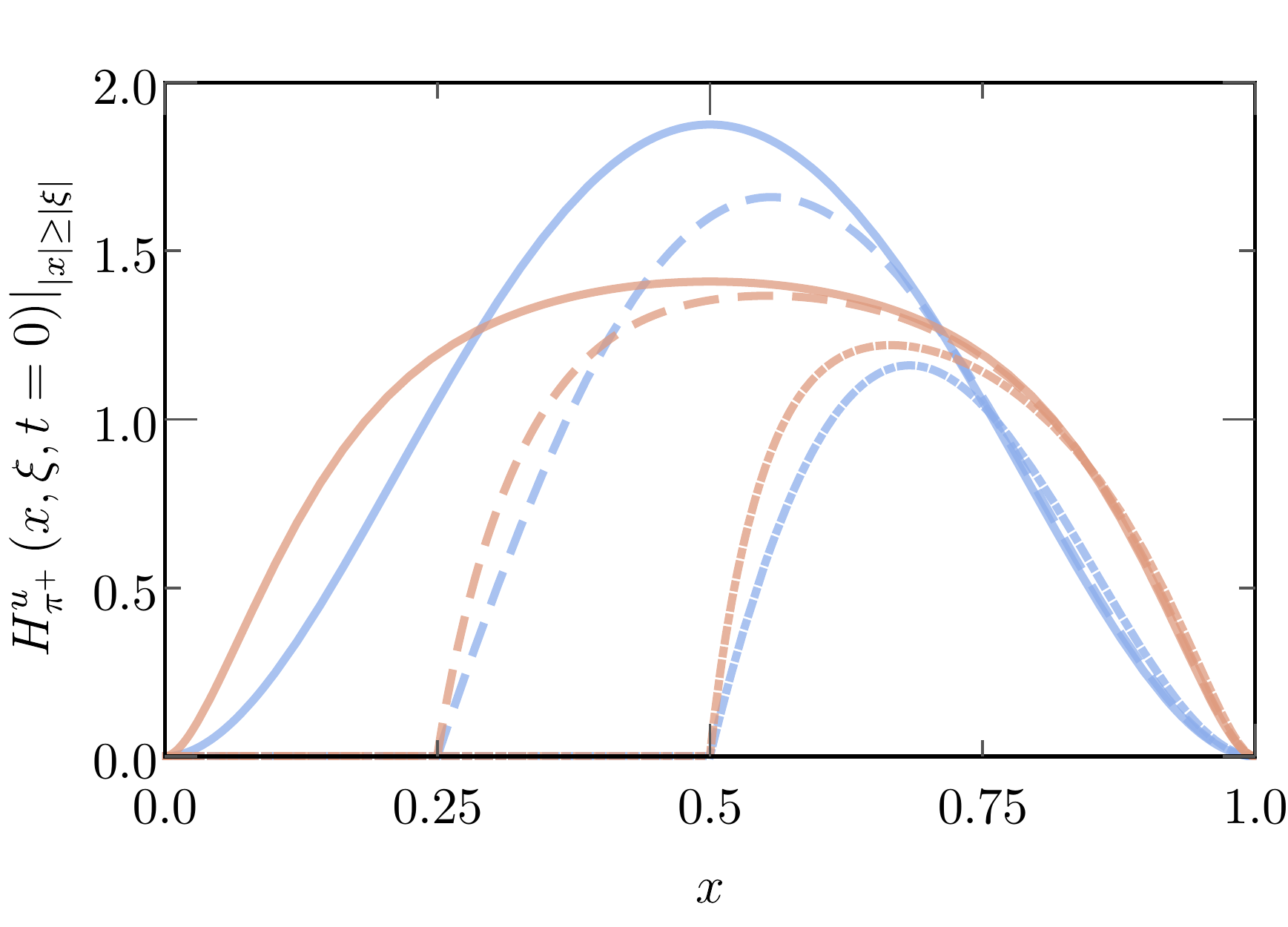}
  \caption{Shapes of the algebraic (blue lines) and numerical (brown lines) pion $u$-quark DGLAP GPDs drawn for: $\xi=0$ (solid), $1/4$ (dashed) and $1/2$ (dot-dashed) and zero momentum transfer.}
  \label{fig:gpdComparison}
\end{figure}

Our approach can be extended to the PDF \eqref{eq:NumericalPDF}. The resulting GPD for $0<\xi<x$ is shown in Fig. \ref{fig:gpdComparison} together with that of Eq. \eqref{eq:AlgebraicModelDGLAP}. Two interesting features are revealed: first, both GPD models are manifestly positive within the DGLAP region; next, the shape of the PDF is ``transferred'' to that of the GPD in the outer subdomain. In that sense, hardening of the GPD's shape within the DGLAP region can be associated to dynamical chiral symmetry breaking, in the sense of Ref. \cite{Chang:2013pq}.

Interestingly, our models are zero at $x=\xi$. This characteristic behaviour arises as a consequence of the end-point behaviour of the factorised LFWFs employed for their development. The $x=\xi$ line gives access to a very particular kinematic configuration: where the momentum fraction along the light-cone carried by the probed parton in the initial hadron state vanishes. Since the two-body leading-twist LFWF reduces to a two-body leading-twist PDA when integrated over $k_{\perp}$, it consistently vanishes at the end-points \cite{Diehl:2003ny}; in agreement with the latter. It also allows us to avoid mathematical complications \cite{Collins:2018aqt}.

\section{Extension to the ERBL region}
\label{sec:Extension}

Following the pioneering work in \cite{Hwang:2007tb}, it was shown in \cite{Chouika:2017dhe} and confirmed in \cite{Muller:2017wms} that, from a GPD model given in the DGLAP subdomain, it is possible to reconstruct uniquely the ERBL region up to D-term like contributions. In this section we sketch the approach followed herein, which relies on a FEM-like (Finite Element Method) strategy to approximate a DD, solve the inverse Radon transform problem to compute such a DD and employ it to ``extend'' a GPD from the DGLAP to the ERBL region. We highlight the improvements performed since \cite{Chouika:2017dhe}, where it was first presented.

\subsection{The covariant extension strategy}

\subsubsection{Double distribution schemes}
\label{subsec:schemes}

As mentioned in Sec. \ref{sec:DD}, DDs are ``scheme dependent'' in the sense that they are not uniquely defined through solely a GPD (see \cite{Teryaev:2001qm,Tiburzi:2004qr,Chouika:2017dhe}). Uniqueness is recovered only when a ``scheme'' is fixed. The GPD itself remains independent of the scheme chosen. In the present paper, we choose to work within the so-called P-scheme \cite{Pobylitsa:2002vi} presented in Eqs. \eqref{eq:PschemeF} and \eqref{eq:PschemeG}. There are several reasons for that. First, this representation was designed in a way which makes it suitable for DD models fulfilling the positivity property. Next, the exact solution to the inverse Radon problem is known for the algebraic model in this specific scheme \cite{Chouika:2017rzs}, providing a natural way to benchmark our code. It also requires only to invert a single function $h_p$ rather than two DDs, $F$ and $G$; while not introducing additional singularities \cite{Chouika:2017dhe}. We will therefore focus in the following on extracting the function $h_{p}$.

\subsubsection{Discretisation and sampling}
\label{subsec:Discretization}

Discretisation of the support domain $\Omega$ of DDs introduced in Sec. \ref{sec:DD} is the first step to carry out before being able to approximate DDs through FEM. This can be done efficiently by taking symmetries into account. The parity in $\alpha$ (see Sec. \ref{sec:DD}) tells us that we can restrict ourselves to the upper half of the $(\beta, \alpha)$ plane. Moreover, because of the structure of the Radon transform in Eq. \eqref{eq:DDDef}, the areas $\beta > 0$ and $\beta < 0$ are probed respectively by ``DGLAP lines'', \emph{i.e.} lines obeying the $x_{i}-\beta-\alpha\xi_{i}=0$ equation such that $|x_{i}| \geq |\xi_{i}|$, with the cases $x>0$ and $x<0$ not mixing among each other. Since the models described in Sec. \ref{subsec:PionGPDs} are identically zero in the negative-$x$ DGLAP region, one can restrict the study without loss of generality to the triangle $\Omega^{+}=\left\lbrace \beta\geq 0\right\rbrace\cap\left\lbrace\alpha\geq 0\right\rbrace\cap\Omega$.

\begin{figure}[b]
  \centering
  \includegraphics[width=0.4\textwidth]{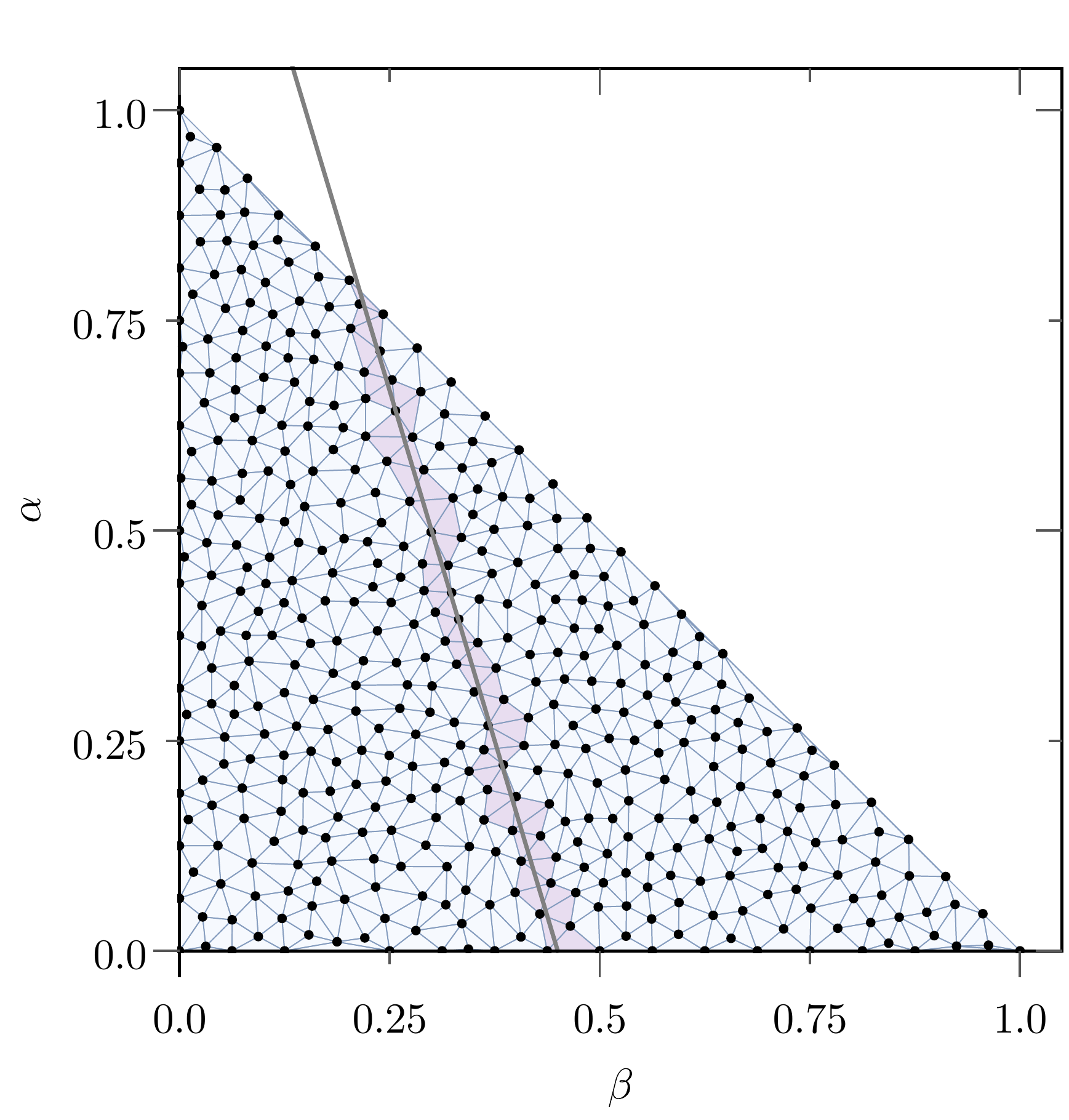}
  \caption{Delaunay triangulation of the upper-right triangle of the DD support, $\Omega^{+}$, with a example ``DGLAP line'' hitting some of the cells. All elements can be sampled by such kind of lines cutting the $\beta=0$ axis outside of the DD support.}
  \label{fig:DelaunayTrig}
\end{figure}

This triangle is then divided in cells through a Delaunay triangulation (chosen for its property of avoiding sliver triangles). We impose a constraint on the maximal possible area of a single element (see Fig. \ref{fig:DelaunayTrig}). With such constraint (set through this work as $0.001$), the \texttt{triangle} discretisation algorithm \cite{triangle} yields a given Delaunay triangulation of the DD support. We end up generating a mesh with $427$ vertices (taken as interpolation nodes) and $780$ elements. After triangulation, the Radon transform problem of Eq. \eqref{eq:DDDef} is reduced to a discrete (matrix) version in the sense of FEM.

Once the $\Omega^{+}$ domain is discretised, it is embedded with a set of basis-functions that allows us to approximate the DD. In this work, these are chosen as two-dimensional degree-one Lagrange polynomials with restricted domain: P1-polynomials. Then each node is allocated with one of such basis-functions so that three conditions are met: (i) the basis-function takes the value one at the corresponding node and zero for all others; (ii) its support spreads over the elements adjacent to such node and (iii) it vanishes on its external boundaries. In this way, all three defining-parameters for each basis-function are unambiguously fixed and continuity of the DD on the cell edges is ensured (see Ref. \cite{Chouika:2017dhe} for further details).

Finally, since the Radon transform operator of Eq. \eqref{eq:DDDef} can be understood as an integral over lines satisfying $x_{i}-\beta-\alpha\xi_{i}=0$, we can choose pairs  $\left(x_{i},\xi_{i}\right)$ within the DGLAP region, build ``DGLAP lines'' and thus sample the DD domain (Fig. \ref{fig:DelaunayTrig}). Through such a sampling process, the Radon transform operator is discretised and we can build a system of linear equations that faithfully represents the corresponding Radon transform problem:
\begin{equation}
  \label{eq:DiscreteRadonTransform}
	B\equiv  
  \begin{pmatrix}
    \\
    b_{i}\\
    \\
  \end{pmatrix}
  =
  \begin{pmatrix}
    & & \\
    & \mathcal{R}_{ij} & \\
    & & 
  \end{pmatrix}
  \begin{pmatrix}
    \\
    d_j\\
    \\
  \end{pmatrix}
\end{equation}
where $d_{j}$ is the unknown value of the DD (represented by a vector D) at a node, $j$, of the mesh. $\mathcal{R}$ is the Radon transform matrix, whose $\left(i,j\right)$ element represents the contribution of the $j$-th basis-function to the approximation of the DD; and $b_{i}=H\left(x_{i},\xi_{i}\right)$ are the elements of the vector, $B$, representing the GPD that must be yielded.

Accessing the ERBL part of a GPD then requires solving the above problem: its solution provides us with the DD associated to the input DGLAP GPD through a Radon transform. Unfortunately, solving such a system of equations may not be straightforward, the reason for that being the \emph{ill-posedness} \cite{Hadamard, Hadamard2, Natterer} of the inverse Radon transform problem.
	
For a better understanding, let us delve into the structure of the Radon transform operator. It has been briefly mentioned that, after discretisation of $\Omega^{+}$, its sampling with DGLAP lines allows us to build a discrete version of such operator: the Radon transform matrix, $\mathcal{R}$. Therefore, accurately sampling such a domain is a crucial step in solving the inverse Radon transform problem. A first possibility in this respect is to choose a number of sampling lines that equals that of nodes. Thus one ends up with a squared $\mathcal{R}$-matrix. However, the geometry of the DD's domain may prevent that system from having a (unique) solution. In a nutshell, the discretisation mesh has most of its nodes located within the low-$\beta$ region (Fig. \ref{fig:DelaunayTrig}). For this reason, naively sampling that domain with the minimum possible number of randomly distributed lines, may result in an under-constrained  system of equations.

An alternative procedure consists in choosing the sampling lines in a smart way, \emph{i.e.} such that every node on the mesh is probed by a given line. Then, the entire discrete-domain can be covered, but a new drawback arises as a consequence of the DD-domain's geometry: in order to probe the low-$\beta$ region, sampling lines with growing slope are needed (Fig. \ref{fig:DelaunayTrig}); thus, because most of the interpolation-nodes are located therein, such procedure ends up with most of the sampling lines being nearly parallel. This brings mostly redundant information that would be numerically compatible with infinitely many solutions.

It is now plain that, for the discrete version of the inverse Radon transform problem, neither the existence nor the uniqueness of a solution can be granted. Both of them sufficient conditions for a problem to be \emph{ill-posed} in the sense of Hadamard. Furthermore, the inverse Radon transform operator (even without discretisation) is non-continuous and thus the stability property of well-posed problems is also violated.

Nevertheless, Lorentz covariance guarantees that physical GPDs are the Radon transform of DDs (see Sec. \ref{sec:DD}), meaning that a unique solution to the inverse Radon problem must exist. It is the discretisation step which may push the GPDs outside of the range of the Radon transform. In fact, this feature is well-known in the context of computerised tomography, where the Radon transform is a common tool, and is referred to as an \emph{inconsistent data problem} \cite{Natterer}. Therefore, facing the problem in Eq. \eqref{eq:DiscreteRadonTransform} requires working out the question of sampling to bypass the inconsistency of the problem at hand.

A possible strategy to deal with such kinds of problems is to build an over-constrained system of equations. Namely, for a sufficiently large number of sampling lines, probing each node can be guaranteed; and, at the same time, the ``noise'' introduced by nearly-parallel sampling lines (low-$\beta$ region) in the solution (DD), attenuated. Therefore, employing a number of sampling lines larger than that of nodes, may allow to overcome the difficulties introduced by the ill-posedness of the inverse Radon transform problem.

\subsubsection{Normal equations}

Provided that the number of sampling lines is large enough, the resulting system of equations is over-constrained and the system's matrix, $\mathcal{R}$, rectangular. In this context, an efficient way to ensure the existence and uniqueness of the solution is to turn to a least-squares formulation:
\begin{equation}\label{eq:leastsquares}
\chi^{2}= \frac{1}{\sigma_{\text{\scriptsize{DGLAP}}}^2} \sum_{i}\left(b_{i}-\mathcal{R}_{ij}d_{j}\right)^{2}
\end{equation}
\emph{i.e.} to search for the DD's values at the interpolation nodes, $d_{j}$, such that the residual $\chi^{2}$ is minimised. Note that we have included a constant uncertainty $\sigma_{\text{\scriptsize{DGLAP}}}$ over the values of the DGLAP region GPD, $b_i$. Notice that, being constant, the uncertainty factor does not impact on the minimisation.

In previous studies of the inverse Radon transform \cite{Chouika:2017dhe}, the solution to the problem of Eq. \eqref{eq:leastsquares} was found by a iterative least-squares algorithm optimised for sparse matrices: the LSMR \cite{LSMR}. In such a context, the residual $\chi^{2}$ is recursively minimised up to a given tolerance, and thus the solution $d_{j}$ obtained.

However, in this work we choose an alternative approach which consists in an exact solution of the optimisation problem of Eq. \eqref{eq:leastsquares}. Minimisation of the residual with respect to $d_{k}$ readily yields the solution to such problem to be given by:
\begin{equation}\label{eq:NormalEqs}
\mathcal{R}^{T}\mathcal{R}D=\mathcal{R}^{T}B\ 
\end{equation}
corresponding to the so-called normal equations of the linear system \eqref{eq:DiscreteRadonTransform}, whose solution provides us with a DD such that $\chi^{2}$ in Eq. \eqref{eq:leastsquares} is minimised. Note that the system of equations obtained here is the same that the one in \eqref{eq:DiscreteRadonTransform} multiplied by the transposed Radon transform matrix, $\mathcal{R}^{T}$, but the system is now squared with the size of the matrix $\mathcal{R}^{T}\mathcal{R}$ being the number of nodes.

For the system above to have a solution, the new system's matrix ($\mathcal{R}^{T}\mathcal{R}$) must be full-rank. Such a condition is satisfied provided that the Radon transform matrix $\mathcal{R}$ has maximal rank (see appendix \ref{app:Invertibility}). Then, the inverse matrix $\left(\mathcal{R}^{T}\mathcal{R}\right)^{-1}$ is proved to exist, and thus the DD which solves the least-squares problem in Eq. \eqref{eq:leastsquares} obtained as:
\begin{equation}
\label{eq:NormalEqsII}
D=\left(\mathcal{R}^{T}\mathcal{R}\right)^{-1}\mathcal{R}^{T}B
\end{equation}
Crucially, for a large enough number of sampling lines, the rank of the Radon transform matrix can be always made maximal. This can be easily understood from the pictorial representation drawn in Fig. \ref{fig:DelaunayTrig}: DGLAP lines can be iteratively added to build the matrix up to the point where $Rank\left(\mathcal{R}\right)=n$, with $n$ the number of nodes making up the mesh. This procedure guarantees that the rank of the matrix $\mathcal{R}$ can be made maximal.

Since the number of sampling lines is a user-defined parameter, we can always assume the involved Radon transform matrix to have maximal rank, and under such assumption solve the inverse Radon transform as a least-squares problem whose solution is given by Eq. \eqref{eq:NormalEqsII}. In fact, once the rank of this matrix is maximal, adding more lines does not modify the system's size but produce larger diagonal elements and hence smaller uncertainties as the covariance matrix is proportional to $(\mathcal{R}^{T}\mathcal{R})^{-1}$ (see Sec. \ref{subsec:Uncertainties}). For this reason, the present method proved to yield more accurate results than the previously used LSMR method. Furthermore, since matrix-inversion routines are, generally speaking, carefully optimised, the normal equations strategy also showed to be much more efficient. Therefore, it was adopted for the covariant extension developed within this work.

\subsubsection{Uncertainty assessment}
\label{subsec:Uncertainties}

With the DD obtained through the inverse Radon transform strategy, the ERBL domain can be accessed; one just needs to sample the $\Omega^{+}$ domain with ``ERBL lines'', \emph{i.e.} choosing pairs $\left(x_{i},\xi_{i}\right)\in\left[-1,1\right]\otimes\left[-1,1\right]\cap\left\lbrace\left|x_{i}\right|\leq\left|\xi_{i}\right|\right\rbrace$, to build the corresponding matrix ($\mathcal{R}_{\text{\scriptsize{ERBL}}}$) and employ it in combination with the DD to evaluate the ERBL GPD at a given point through:
\begin{equation}
\label{eq:DiscreteCovExt} 
B_{\text{\scriptsize{ERBL}}}=\mathcal{R}_{\text{\scriptsize{ERBL}}}D
\end{equation}

One further virtue of the normal equations strategy is that it provides a direct and clear window onto the assessment of the uncertainties originated by discretisation and interpolation of the DD. 
Indeed, when solving the least-squares problem derived from \eqref{eq:leastsquares}, the covariance ($C$) matrix of the solutions $D$, is given by $C=\sigma_{\text{\scriptsize{DGLAP}}}^2\left(\mathcal{R}^{T}\mathcal{R}\right)^{-1}$. And solving the inverse Radon transform problem through Eq. \eqref{eq:NormalEqsII} requires the knowledge of the matrix $\left(\mathcal{R}^{T}\mathcal{R}\right)^{-1}$.

The uncertainty of the results of the GPD's covariant extension to the ERBL region can be obtained applying standard uncertainty propagation to \eqref{eq:DiscreteCovExt}, and is given by \cite{NumRec}:
\begin{eqnarray}
\sigma^2_{\text{\scriptsize{ERBL}},\ i} &=& \sum_{jk} \frac{\partial B_{\text{\scriptsize{ERBL}},\ i}}{\partial d_j} \frac{\partial B_{\text{\scriptsize{ERBL}},\ i}}{\partial d_k} C_{jk} \\ 
&=& \sigma_{\text{\scriptsize{DGLAP}}}^2 \left(\mathcal{R}_{\text{\scriptsize{ERBL}}}
\left(\mathcal{R}^{T}\mathcal{R}\right)^{-1}
\mathcal{R}_{\text{\scriptsize{ERBL}}}^t\right)_{ii} \nonumber
\end{eqnarray}

Therefore, the only ingredient that remains to be estimated is the uncertainty $\sigma_{\text{\scriptsize{DGLAP}}}$ associated to the DGLAP GPD yielded by our numerically computed DD. Here, we adopt a conservative approach and estimate it as the maximum separation between the input and numerical DGLAP GPDs, $\sigma^2_{\text{\scriptsize{DGLAP}}}=\max_i \left(b_i-\sum_j \mathcal{R}_{ij} d_j\right)^2$. Then, the covariance and $\mathcal{R}_{\text{\scriptsize{ERBL}}}$ matrices allows us to propagate such an uncertainty to ERBL region.

\subsubsection{Soft-pion theorem}
\label{subsec:softpionth}

After the covariant extension explained above, one is still left with the issue of the D-term ambiguities, as explained in Refs. \cite{Chouika:2017dhe,Muller:2015vha}. In order to handle them, we exploit the soft-pion theorem mentioned in Eq. \eqref{eq:SoftPionTheorem}. Technically speaking, both $x$-even and $x$-odd ambiguities can arise, but since the even one does not play any role in the computation of the CFFs, we will focus on the odd one, that we call the extrinsic D-term. The interested reader can find more details in \cite{Chouika:2017dhe,Chouika:2017rzs} and references cited therein. 

At vanishing momentum transfer, Eq. \eqref{eq:SoftPionTheorem} tells us that the quark GPD needs to be even at all scales, \emph{i.e.} the singlet and gluon GPDs are vanishing in the limit $(\xi,t) \to (1,0)$. This condition naturally allows us to fix the extrinsic D-term $D^{q/g}$ at all scales following:
\begin{equation}
  \label{eq:DtermQuarks}
  D^q(z,0;\mu^2) = -\frac{H^q(z,1,0;\mu^2)-H^q(-z,1,0;\mu^2)}{2}
\end{equation}
for quarks and,
\begin{equation}
  \label{eq:DtermGluons}
    D^g(z,0;\mu^2) = -\frac{H^g(z,1,0;\mu^2)+H^g(-z,1,0;\mu^2)}{2}
\end{equation}
for gluons.

However, the soft-pion theorem does not provide any information about the $t$-dependence of the D-term. To constrain the latter, we rely on pQCD predictions, stating that at large $-t$, the moments of the pion's GPD behave like $1/|t|$ up to logarithmic corrections \cite{Hoodbhoy:2003uu}. We therefore chose a monopole description using the same mass scale as the one previously advocated:
\begin{equation}
  \label{eq:DtermtDependence}
  D^{q/g}(z,t;\mu^2) = \frac{D^{q/g}(z,0,\mu^2)}{1-\frac{t}{M^2}}
\end{equation}
with $M$ already introduced in Eq. \eqref{eq:zetaDef}, and used it for all quark flavours and gluons.

\subsubsection{Validation}

\begin{figure}[b]
\centering
\includegraphics[scale=0.6]{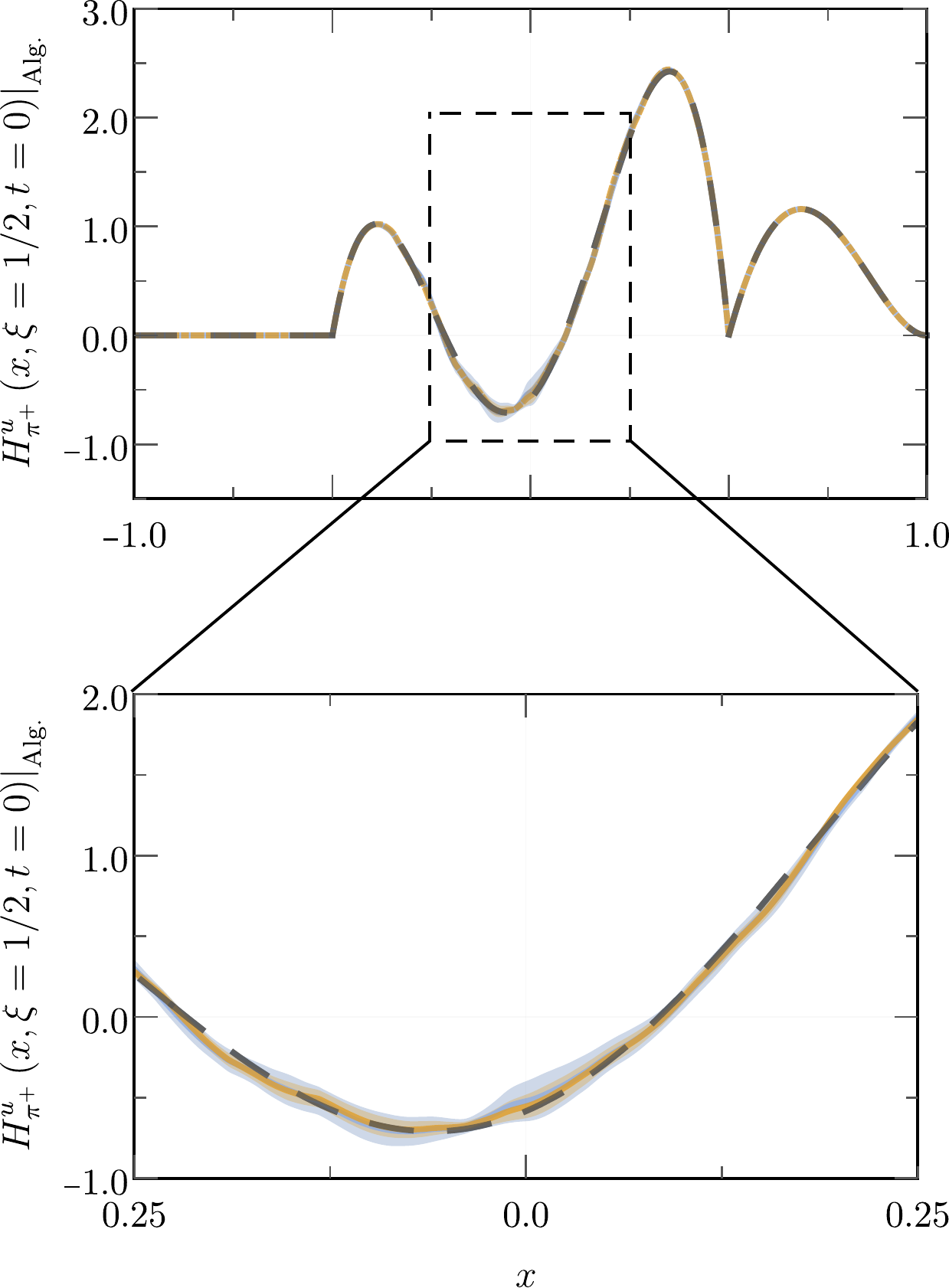}
\caption{Algebraic model pion GPD at $\xi=1/2$ and $t=0$. Comparison between the exact, analytical result (dashed black line) \cite{Chouika:2017rzs} and two solutions of the associated numerical problem: $\mathcal{R}$ built with $3120$ (blue) and $9360$ (orange) sampling lines.}
\label{fig:BenchmarkingAlgModel}
\end{figure}

Once we presented the general idea behind the covariant extension strategy, we can exploit it for the extension of a given model and thus illustrate its advantages and drawbacks. For this purpose, since its ERBL region is known in closed form \cite{Chouika:2017rzs}, the algebraic model pion DGLAP GPD discussed in Sec. \ref{subsec:PionGPDs} constitutes an outstanding benchmark, allowing for comparison with the numerical results obtained through the approach presented in the previous sections.

As an illustration, Fig. \ref{fig:BenchmarkingAlgModel} shows a comparison between the results obtained through the numerical implementation of the covariant extension strategy described before (blue and orange lines) and the actual, analytical result \cite{Chouika:2017rzs} (dashed-black) at $\xi=1/2$ and $t=0$. 

For this example, the Radon transform matrix was filled by sampling the $\Omega^{+}$ domain with $3120$ ``DGLAP lines'' (corresponding to four times the number of mesh cells, $n_{e}$), represented in blue; and 9360 sampling lines ($12n_{e}$), in orange. Both configurations proved to yield $\mathcal{R}$-matrices of maximal rank. Then the system of equations was solved by means of the normal equations strategy. Together with the numerical solutions, the corresponding error bands are shown (one standard deviation).

From Fig. \ref{fig:BenchmarkingAlgModel} it is plain that the numerical approach described above yields satisfactory results, the numerical solution being essentially indistinguishable from the analytical result. Only in the inner region deviations from the actual curve can be observed, but even there, the analytical curve is always lying within the corresponding uncertainty band for both numerical solutions. This agreement was verified to remain true for calculations with different configurations (number of sampling lines) and kinematic configurations. From a statistical perspective, we would have expected the analytical results to be sometimes out of the error band. As it is not the case, we deduce that our choice for $\sigma_{\text{\scriptsize{DGLAP}}}$ is probably too conservative, generating uncertainty bands which are too large. We nevertheless stick to that choice in the following to assess an order of magnitude of the uncertainties generated by the numerical inversion.

Finally, both configurations of the sampling strategy proved to yield nearly identical results, the uncertainty band associated to the solution using a larger number of sampling lines turned out to be narrower at the price of increasing the computation time. This finding was in fact expected as increasing the number of sampling lines allowed us to build system's matrices with larger eigenvalues, thus favouring a better performance of the inversion routine for $\mathcal{R}^{T}\mathcal{R}$. An exhaustive analysis confirmed this observation and showed that the configuration with $12n_{e}$ randomly-distributed sampling lines, which we employ for the rest of this work, represents the optimal compromise between accuracy and performance.

\subsection{Covariant extension of the numerical model}

\subsubsection{Numerical inversion}
  
Once the formalism of the covariant extension has been presented and validated, we apply it for the completion of our two GPD models (Sec. \ref{subsec:PionGPDs}). The algebraic one is continued to the ERBL domain following the approach of \cite{Chouika:2017rzs}, \emph{i.e.} exactly solving the inverse Radon transform problem. On the other hand, the DGLAP GPD model built from a the pion-PDF parametrisation of Eq. \eqref{eq:NumericalPDF} is extended by means of the numerical procedure developed above.

\begin{figure}[t]
  \centering
  \includegraphics[scale=0.45]{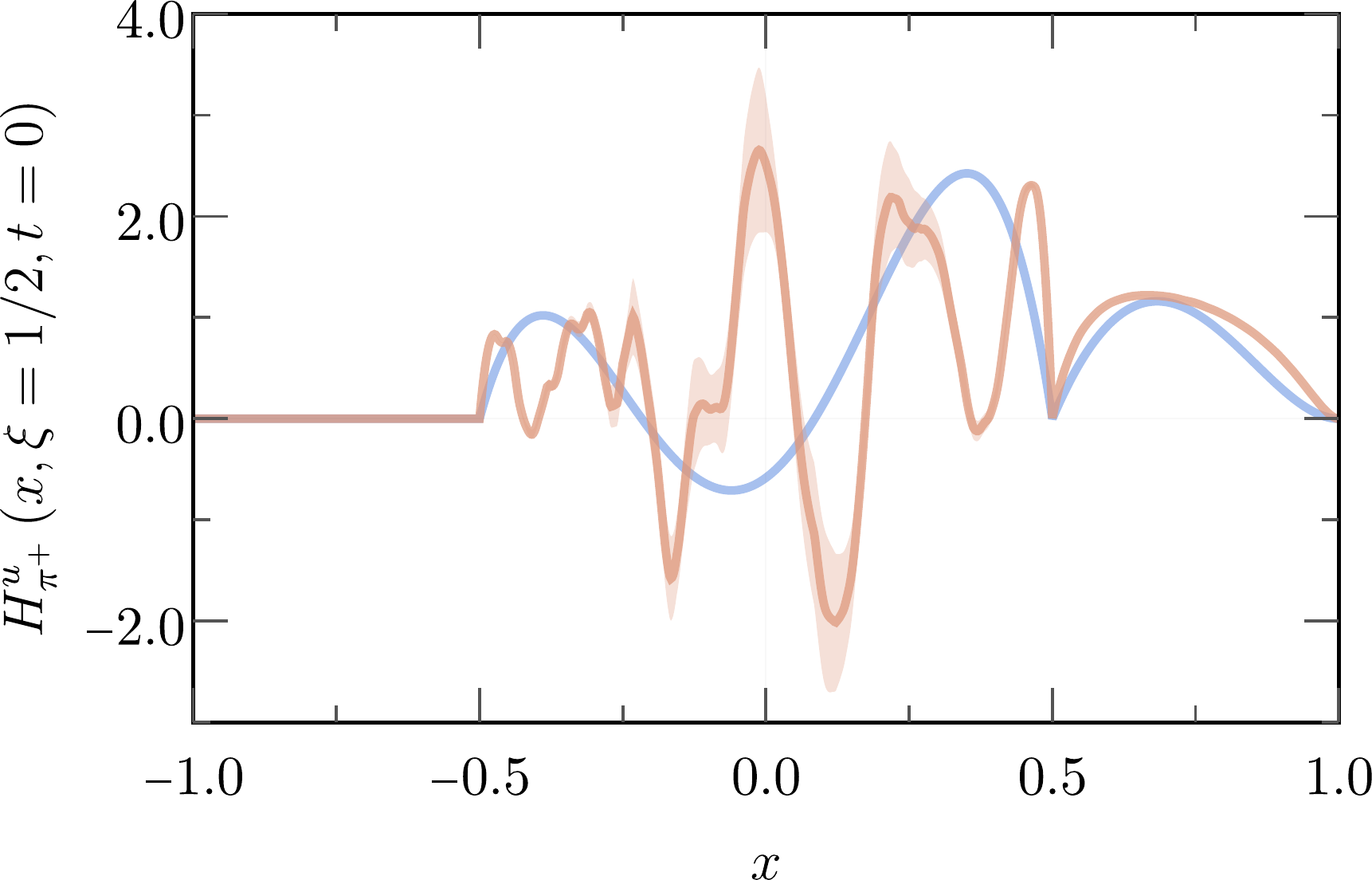}
  \caption{Algebraic and realistic models evaluated at vanishing momentum transfer for $\xi = 1/2$ after fixing the D-term ambiguity with the soft-pion theorem.}
  \label{fig:CSMGPDmodels}
\end{figure}

As an illustration, Fig. \ref{fig:CSMGPDmodels} shows the resulting GPDs at vanishing $t$ and $\xi=1/2$. The numerical model exhibits an oscillating behaviour within the ERBL region, more marked than the one of the algebraic model. Such a behaviour is confirmed by the conservative assessment of the error band associated to our numerical extraction. Indeed, the error-band is large at the top of the oscillation, but small around the zero crossings, confirming the oscillating pattern shown by the numerical GPD.

Beyond this oscillating behaviour, we stress that the continuity at the crossover lines $|x|=|\xi|$, highlighted in Sec. \ref{subsec:Definition and Properties} (see also Ref. \cite{Collins:1998be}), is indeed preserved throughout our work; despite the inverse Radon transform not being continuous itself. Interestingly, following other types of model based on DDs (see \emph{e.g.} \cite{Musatov:1999xp,Mezrag:2013mya}), the first derivative is not continuous. This ``singularity'' is inherited from the behaviour of the DDs on the corners of their definition domain \cite{Tiburzi:2004qr}, and is consistent with the LO evolution kernel \cite{EvolutionPaper}.

Last but not least, let us mention that this continuity property is a key point for being able to describe exclusive processes, whose factorisation theorem is inconsistent with discontinuous GPDs on the crossover lines. It thus makes the pion GPD models developed here suitable for phenomenological applications, guaranteeing the calculation of CFFs to yield finite results (see Sec. \ref{sec:CFF}).

\subsubsection{Electromagnetic and gravitational form factors}

Once the GPD models are defined over the entire kinematic domain, they can be exploited for the calculation of Mellin moments, in general, and electromagnetic or gravitational form factors, in particular.
	
\begin{figure}[t]
\centering
\includegraphics[scale=0.45]{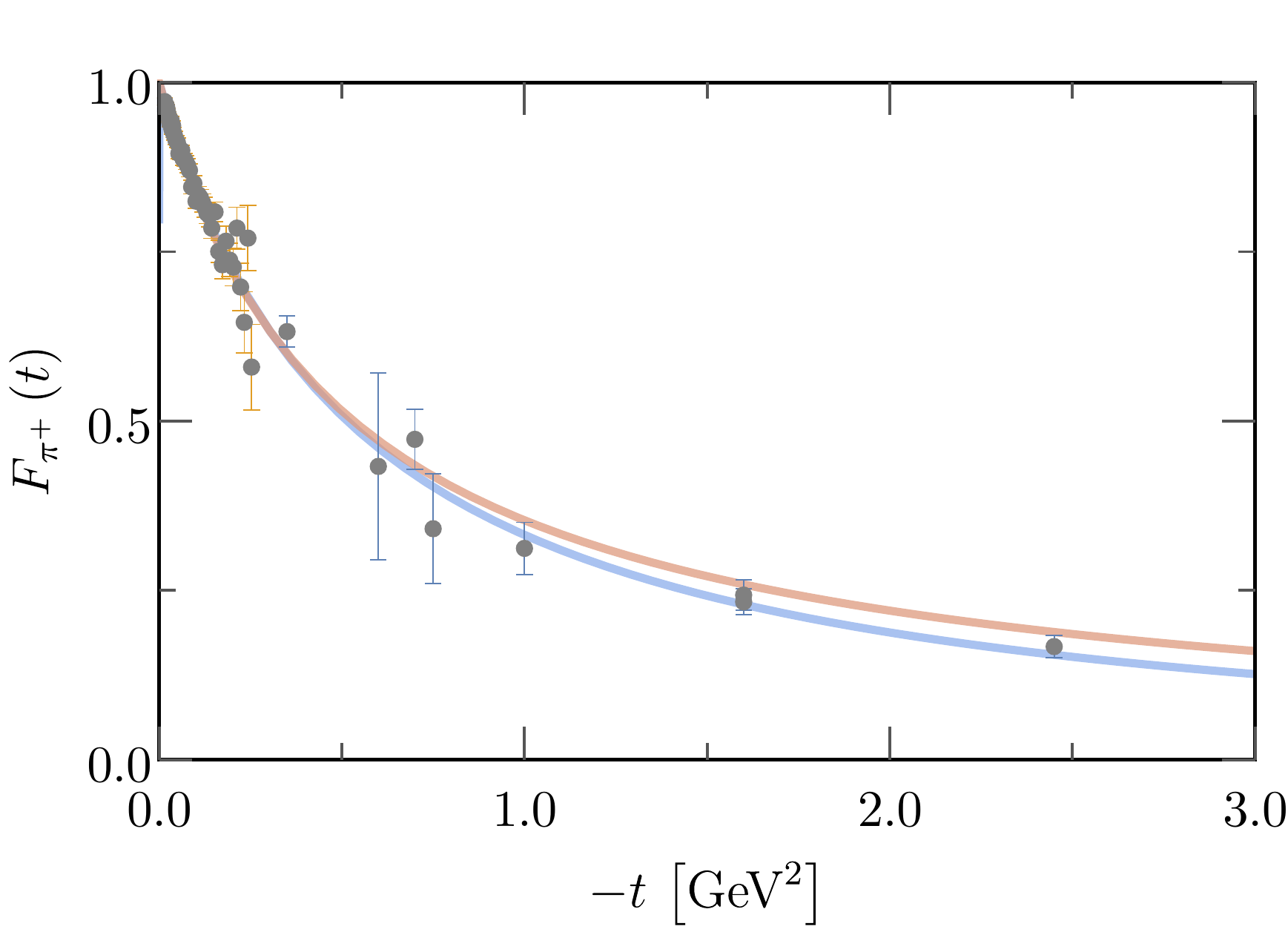}
\caption{Calculation of pion's electromagnetic form factors within the two models discussed through this text: algebraic (blue line) and numerical model (brown). Comparison with experimental data from \cite{Huber:2008id, Amendolia:1986wj}.}
\label{fig:EFF}
\end{figure}

Coming back to Eq. \eqref{eq:PolynomialityQuark}, we recall that the pion EFF can be computed for each quark flavour as (Fig. \ref{fig:EFF}):
\begin{equation}
F^q\left(t\right)\equiv A^{q}_{1,0}\left(t\right)=\int_{-1}^{1}dx H^{q}_{\pi}\left(x,\xi,t\right)
\end{equation}

Notice that such a Mellin moment does not depend on the skewness variable and thus can be obtained from a direct integration of GPDs as defined in the forward limit (DGLAP GPD with $\xi=0$). This property allows for a cross-checked calculation of the EFF: both using covariantly extended and forward limit only GPDs. As expected, this approach revealed no dependence of the pion EFF on the ERBL completion strategy.

Results obtained for both algebraic and realistic models show a good agreement with available experimental data \cite{Huber:2008id, Amendolia:1986wj}, especially when $|t| \le 1\,\textrm{GeV}^2$ (Fig. \ref{fig:EFF}); the kinematic region that we are mostly interested in. Moreover, such satisfactory results are obtained with one single free parameter: the mass-scale $M$ arising in the LFWFs (see Sec. \ref{sec:DGLAPRegion}). It was fitted to the pion's electromagnetic charge radius, $r_{\pi}=0.672\pm 0.008\text{ fm}$ \cite{PDG2012} through:
\begin{equation}\label{eq:ChargeRadius}
r_{\pi}^{2}=\left.-6\frac{F_{\pi}\left(-t\right)}{d\left(-t\right)}\right|_{t=0}\Rightarrow F_{\pi}\left(-t\right)\simeq 1 - \frac{r_{\pi}^{2}}{6}\left(-t\right)
\end{equation}
yielding a value of $M=318\,\text{MeV}$ for both models. In this sense, the results shown on Fig. \ref{fig:EFF} can be understood as model predictions for the pion EFF.
Getting this accurate fit of experimental data also reveals a deeper implication: the absence of some components of the pion's BSA (\emph{e.g.} pseudovector components), does not introduce significant deviations from the expected monopole-like behaviour (within the explored $t$-range).

With the models at hand it is possible to get a step forward and compute higher-order Mellin moments. To this end, we recall that we can rewrite the GPDs as:
\begin{align}
  \label{eq:DtermExtInt}
  H\left(x,\xi,t\right)=&H_{D}\left(x,\xi,t\right)\nonumber  \\
  & +\text{sign}\left(\xi\right)\left[D^{\textrm{Int}}\left(\frac{x}{\xi},t\right)+D^{\textrm{Ext}}\left(\frac{x}{\xi},t\right)\right]
\end{align}
where we have restored the explicit reference to the $t$-dependence. $H_{D}\left(x,\xi,t\right)$ denotes the GPD yielding the $A$ generalised form factors in Eq. \eqref{eq:PolynomialityQuark}, while the D-terms generate the $C$ ones. We highlight the fact that $D^{\textrm{Int}}$ is the intrinsic contribution to the D-term, \emph{i.e.} that generated by the DD $G\left(\beta,\alpha,t\right)= -\alpha h_{P}\left(\beta,\alpha,t\right)$ (see Eq. \eqref{eq:DDtoD}); while $D^{\text{Ext}}$ is added as allowed \cite{Muller:2015vha,Chouika:2017dhe} and tuned so that the soft-pion theorem is fulfilled.

\begin{figure*}[t]
  \centering
  \begin{subfigure}{0.49\textwidth}
  \centering
  \includegraphics[scale=0.45]{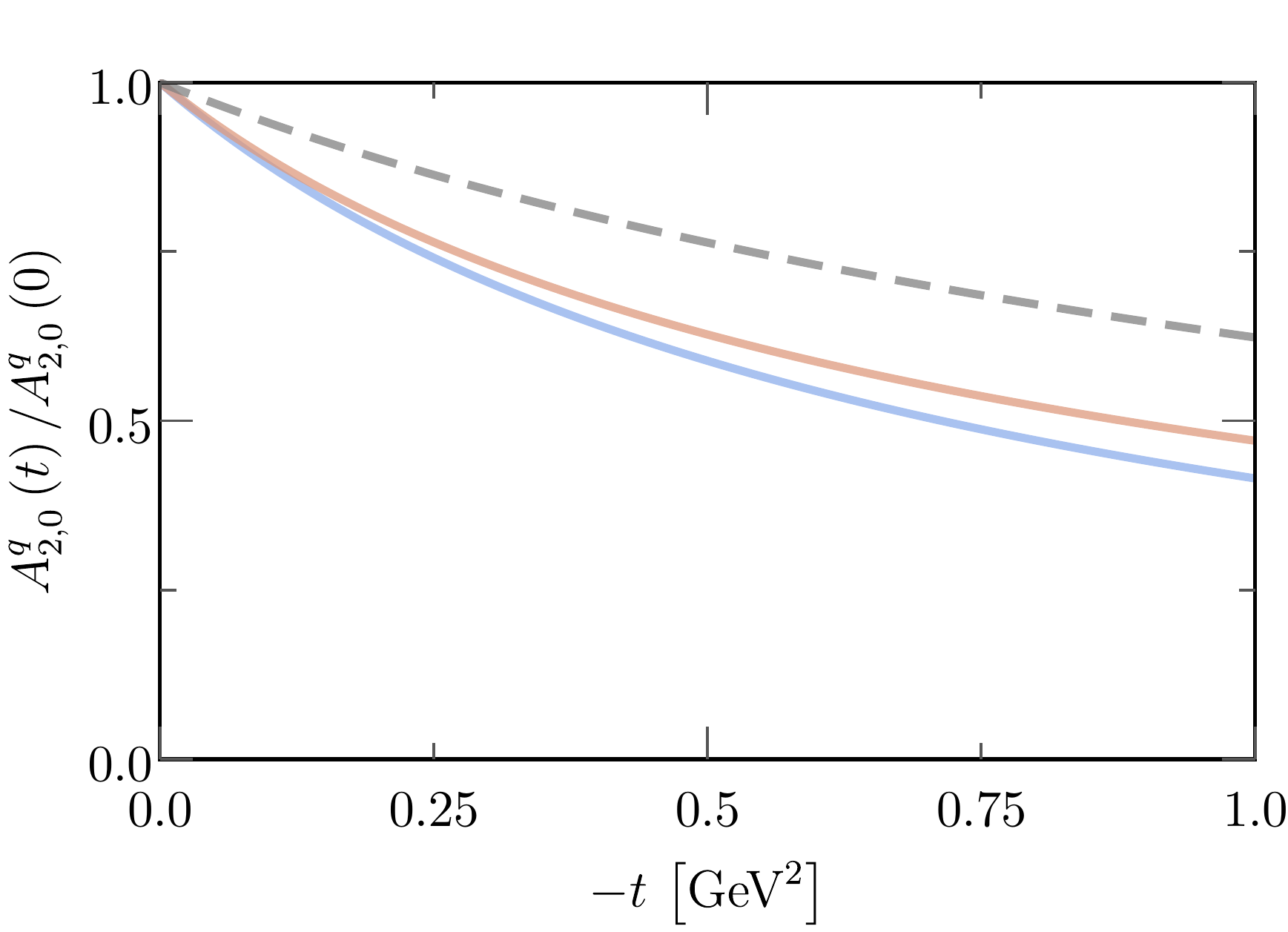}
  \end{subfigure}
  \begin{subfigure}{0.49\textwidth}
  \centering
  \includegraphics[scale=0.45]{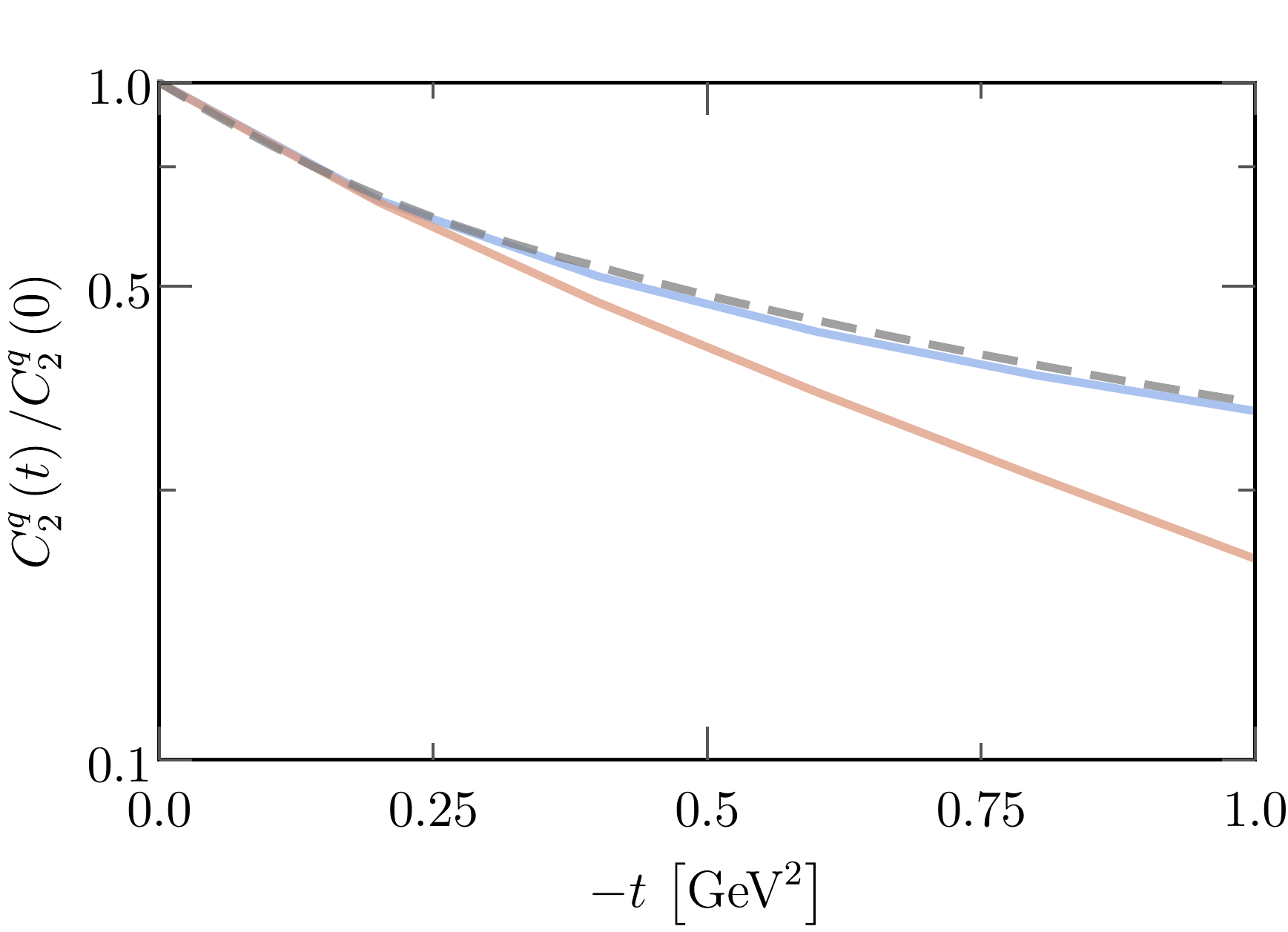}  
  \end{subfigure}
  \caption{\textit{Left panel} - Unit-normalised results for the gravitational form factor $A^{q}_{2,0}\left(-t\right)$ computed through the algebraic (blue line) and numerical models (brown line). \textit{Right panel} - Logarithmic-scale plot of the unit-normalised gravitational form factor $C^{q}_{2}\left(-t\right)$ computed through the algebraic (blue line) and numerical models (brown line). Dashed grey line represents the latest extractions from $\gamma^{\ast}\gamma\to\pi^{0}\pi^{0}$ experimental data \cite{Kumano:2017lhr}.}
  \label{fig:GFFComparison}
\end{figure*}

Then, we compute first-order Mellin moments of GPDs. Following Eq. \eqref{eq:PolynomialityQuark}, it reads:
\begin{equation}
\int_{-1}^1\textrm{d}x\, x H^{q}_{\pi}(x,\xi,t)=A^{q}_{2,0}\left(t\right)+4\xi^{2}C^{q}_{2}\left(t\right)
\end{equation}
where, $A^{q}_{2,0}\left(t\right)$ can again be obtained from $\xi\to 0$ limit GPDs, and therefore presents no special difficulty. The result yielded by our two models is shown in the upper panel of Fig. \ref{fig:GFFComparison}, together with the most recent experimental extraction \cite{Kumano:2017lhr}. There, a faster decay with the momentum transfer (with respect to experimental data) is observed in both cases, meaning that the expected $1/\left|t\right|$ behaviour might be violated due to missing contributions in the BSA (see  sec. \ref{subsec:PionGPDs}), marking a significant difference with the EFF.

On the contrary, $C^{q}_{2}$ is purely generated from the first-order Mellin moment of the sum between the intrinsic and extrinsic D-terms (Eq. \eqref{eq:DtermExtInt}). The soft-pion theorem provides an unambiguous way to fix $D^{\textrm{Ext}}$ at $t=0$, but as discussed before, its $t$-dependence is left unconstrained. It therefore requires an additional modelling assumption. We decide here to describe it as a monopole (see Eq. \eqref{eq:DtermtDependence}). We assume the mass scale therein to be that in the LFWF; which we fixed through the pion's electromagnetic charge radius (as discussed above). Neither this choice of a monopole-like parametrisation nor the existence of one single mass scale rely on first-principles arguments (at least, when working outside Polyakov-Weiss' DDs scheme). However this simple approach revealed in Fig. \ref{fig:GFFComparison} (right panel) ay low $|t|$ (below $0.5-0.6\,\text{GeV}^2$) a consistent behaviour between the two GPD models on the one hand, and between them and existing extractions for such GFFs \cite{Kumano:2017lhr}, on the other hand. This is indeed a crucial requirement for these GPD models which, apart from fulfilling all the requirements imposed by QCD, are intended to be exploited in the assessment of DVCS (see Sec. \ref{sec:Phenomenology}).

DVCS can be described in terms of GPDs in the low-$|t|$ regime \cite{Sullivan:1971kd, Amrath:2008vx}. Therefore the GPDs behaviour within that region deserves special attention. In this respect, the ``pressure-radius'' allows for a fair quantification of our accuracy at small momentum transfer. In fact, $r^{\theta_{1}}_{\pi}$ can be defined analogously to Eq. \eqref{eq:ChargeRadius} \cite{Zhang:2021mtn, Raya:2021zrz}, yielding for our two models:
\begin{equation}
\left.\frac{r_{\pi}^{\theta_{1}}}{r_{\pi}}\right|_{\text{Alg.}}=1.17\qquad \left.\frac{r_{\pi}^{\theta_{1}}}{r_{\pi}}\right|_{\text{Ding}}=1.07
\end{equation}
These results are in agreement with those extracted from $\gamma^{*}\gamma\to\pi^{0}\pi^{0}$ \cite{Kumano:2017lhr}. Despite the existing model dependence and the simple choice for the D-term's momentum transfer dependence, Eq. \eqref{eq:DtermtDependence}, the slope at $t\to 0$ of the pressure distribution matches the expectation, even when fixed through an independent quantity ($r_{\pi}$) and thus supports the choice of a monopole-like \textit{Ansatz} for the D-term when the region of interest is that of low $t$. 

Summarising this section, we obtained GPD models which are unambiguously defined, matching both theoretical and phenomenological expectations, and their suitability for phenomenological analyses becomes manifest.

\section{Phenomenological modelling}
\label{sec:Pheno}

\begin{figure}[t]
\centering
	\begin{subfigure}{0.5\textwidth}
	\centering
	\includegraphics[width=0.9\textwidth]{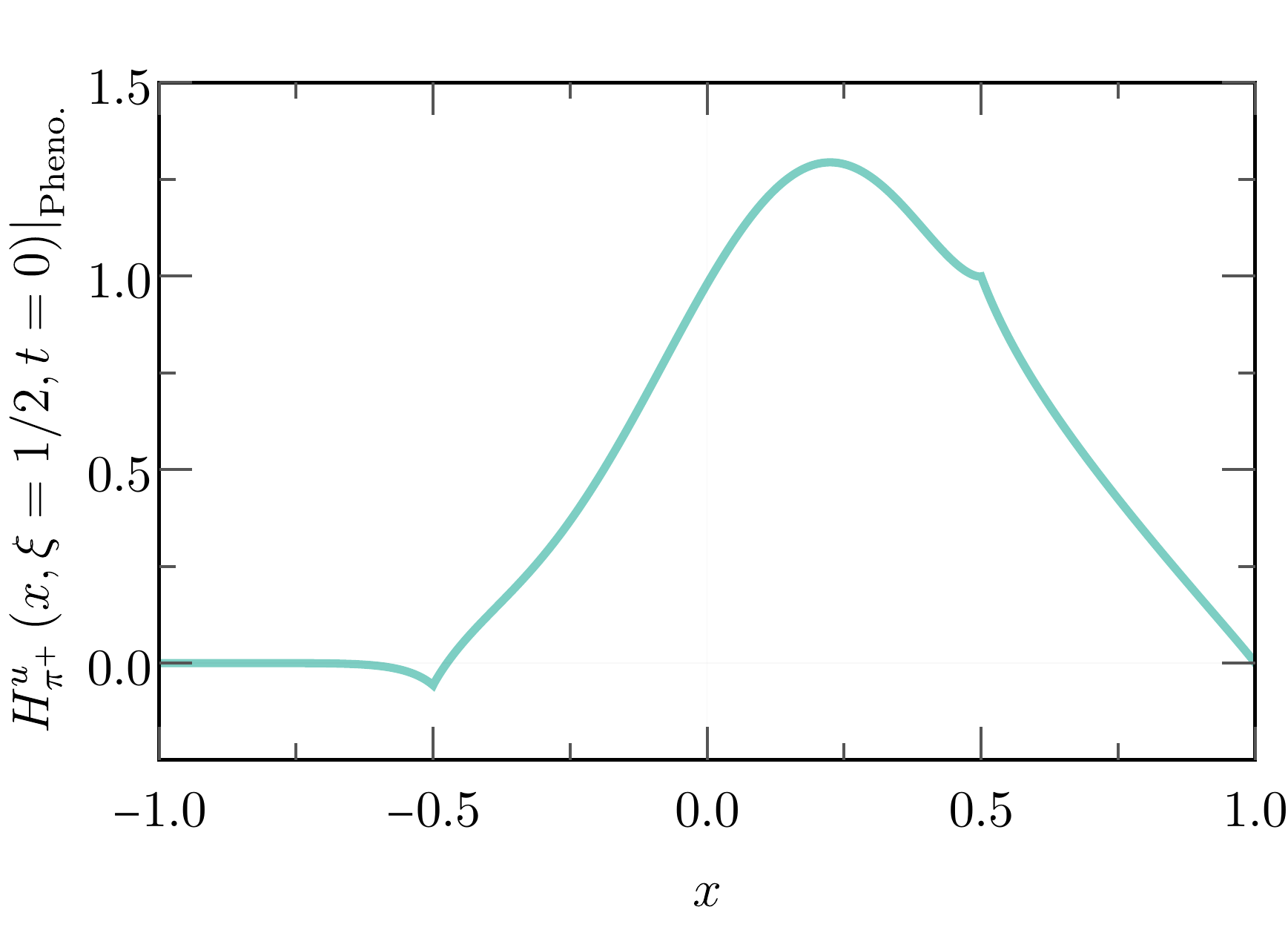}
	\end{subfigure}
	\centering
	\begin{subfigure}{0.5\textwidth}
	\centering
	\includegraphics[width=0.9\textwidth]{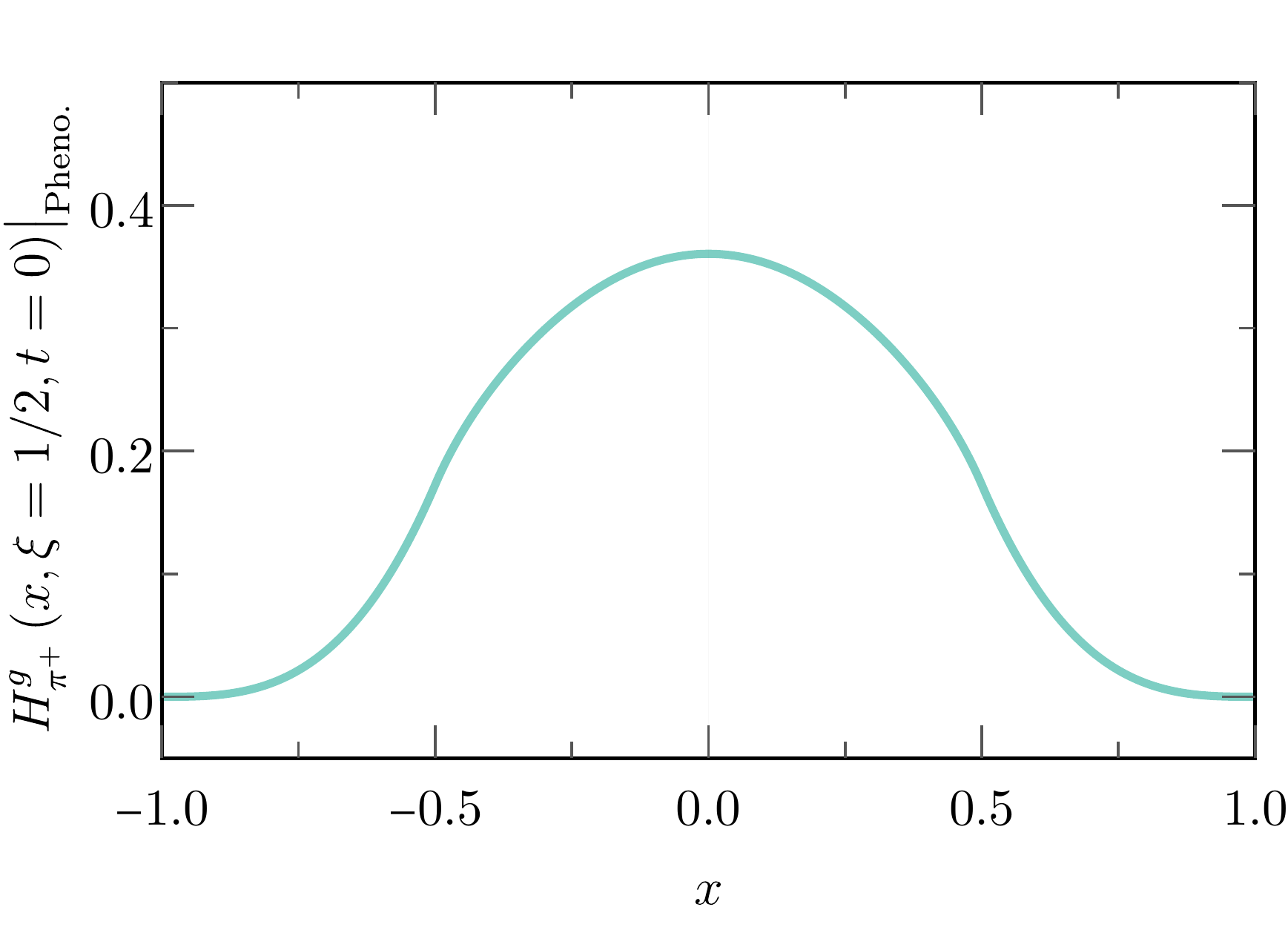}  
	\end{subfigure}
	\caption{\textit{Upper panel} - Phenomenological quark GPD model taken at $\xi=1/2$ and $t=0$. \textit{Lower panel} - Phenomenological gluon GPD evaluated at $\xi=1/2$ and $t=0$. Both shown at the original scale of $\mu^{2}=1.9~\textrm{GeV}^2$.}
	\label{fig:PhenoGPDs}
\end{figure}

In order to get a comparison at the level of CFFs, since until now no GPD related experimental data are available, we introduce a ``phenomenology-like'' model based on the Radyushkin Double Distribution \emph{Ansatz} (RDDA) \cite{Musatov:1999xp}, the xFitter pion PDF set \cite{Novikov:2020snp} and the $t$-dependence suggested in \cite{Amrath:2008vx}. In brief, we model the quark and gluon GPDs as:
\begin{align}
  \label{eq:PhenoGPDq}
  H_{\pi}^q(x,\xi,t) &= \int\textrm{d}\Omega q_\pi(\beta) h(\beta,\alpha) r(\beta,t) + \frac{\xi}{|\xi|} D_q\left(\frac{x}{\xi},t\right),\\
  \label{eq:PhenoGPDg}
    H_{\pi}^g(x,\xi,t) &= \int\textrm{d}\Omega  \beta g_\pi(\beta) h(\beta,\alpha) r(\beta,t) + |\xi| D_g\left(\frac{x}{\xi},t\right),
\end{align}
where $\textrm{d}\Omega = \textrm{d}\beta \textrm{d}\alpha \delta(x-\beta- \alpha \xi) \theta(1-|\beta|-|\alpha|)$. The PDFs are given by:
\begin{align}
  \label{eq:valencepdf}
  x q_v(x) & = \frac{1}{2}A_v x^{B_v}(1-x)^{C_v} \\
  \label{eq:seapdf}
  x q_s(x) &= \frac{1}{6} \frac{A_S}{\mathcal{B}(B_s+1,C_s+1)} x^{B_s}(1-x)^{C_s}\\
  \label{eq:updf}
  u_\pi(x) & = -\theta(-x)q_s(|x|)+\theta(x)\left(q_v(x)+q_s(x) \right)\\
  \label{eq:dpdf}
  d_{\pi}(x) & = -\theta(-x)\left(q_v(|x|)+q_s(|x|)\right)+\theta(x)q_s(x)\\
  \label{eq:spdf}
  s_{\pi}(x) & = -\theta(-x)q_s(|x|)+\theta(x)q_s(x)\\
  \label{eq:gpdf}
  x g_\pi(x) &= A_g(C_g+1)(1-x)^{C_g}
\end{align}
where $\mathcal{B}$ is the Euler beta function and the parameters $A_i,B_i$ and $C_i$ are taken as the central values of the xFitter fit obtained in \cite{Novikov:2020snp}. For completeness they are recalled in Table \ref{tab:PDFparam}.
\begin{table}[b]
\centering
\begin{ruledtabular}
  \begin{tabular}[c]{|c|ccc|}
        & $A_i$ & $B_i$    & $C_i$ \\\hline
$i=v$   & 2.60  & 0.75     & 0.95  \\
$i = s$ & 0.21  & 0.5      & 8     \\
$i=g$   & 0.23  & $\times$ & 3     \\
  \end{tabular}
\end{ruledtabular}
  \caption{Parameters for pion PDFs obtained in \cite{Novikov:2020snp}}
  \label{tab:PDFparam}
\end{table}
The profile function $h(\beta,\alpha)$ is given by the RDDA:
\begin{align}
  \label{eq:ProfileFunction}
  h(\beta,\alpha) = \frac{\Gamma(2N+2)}{2^{2N+1}\Gamma^2(N+1)} \frac{((1-|\beta|)^2-\alpha^2)^N}{(1-|\beta|)^{2N+1}}
\end{align}
where we choose $N=2$. The $r(\beta,t)$ function is then chosen following \cite{Amrath:2008vx,Diehl:2004cx}, \emph{i.e.} based on Regge phenomenology:
\begin{align}
  \label{eq:rfunction}
  r(\beta,t) & = \exp\left(t f(|\beta|) \right), \nonumber \\
  f(\beta) & = (1-x)^3 \left(\kappa \ln \left(\frac{1}{x}\right) + B \right) + A\beta (1-\beta)^2,  
\end{align}
where $\kappa = 0.9~ \textrm{GeV}^2$, $A$ and $B$ being fitted to the values of the EFF. We obtained $A =1.48~\textrm{GeV}^2 $ and $B= 1.14~ \textrm{GeV}^2$, \emph{i.e.} the same order of magnitude than the authors of Ref. \cite{Amrath:2008vx}. We make the rough assumption that these parameters are the \emph{same} for both quarks and gluons, since the absence of gluon sensitive data precludes any sensible fit. Finally, just like before, one is left with determining the values of the D-terms $D^q(\alpha)$ and $D^g(\alpha)$, this time both for quarks and gluons. Once more, we apply the soft-pion theorem, enforcing that:
\begin{align}
  \label{eq:PhenoSoftPion}
  H^q(-x,1,0) = H^q(x,1,0),\quad H^g(x,1,0) = 0 .
\end{align}
We therefore define:
\begin{align}
  \label{eq:D0quark}
  D_0^q(x) & = D^q(x,0) = \frac{H^q(-x,1,0)-H^q(x,1,0)}{2} \\
  \label{eq:D0gluon}
  D_0^g(x) & = D^g(x,0) = -H^g(x,1,0)
\end{align}
Finally, the $t$-dependence of the quark D-term is fitted to the gravitational form factors \cite{Kumano:2017lhr} through the Ansatz:
\begin{align}
  \label{eq:quarkPhenoDtermAnsatz}
  D^q(z,t) = \frac{D_0^q (z)}{1-\frac{t}{\Lambda^2_{D^q}}},
\end{align}
and we found $\Lambda^2_{D^q} = 0.53~ \textrm{GeV}^2$. We apply the same \emph{Ansatz} for the gluon and once again, since there is no data available for the latter, we assume $\Lambda^2_{D^g}=\Lambda^2_{D^q}$. Fig. \ref{fig:PhenoGPDs} shows the typical results obtained for these GPDs. The continuous albeit non differentiable property of the GPD at $|x|=|\xi|$ is again manifest on the figure for quarks, and harder to see but present for gluons.

\section{From GPDs to Compton Form Factors}
\label{sec:Phenomenology}

Now, we would like to assess how the differences between the three models translate into experimental observables that may be reachable through, \emph{e.g.}, the Sullivan process \cite{Amrath:2008vx}. It should be noted that we have not taken into account virtuality effects here, which can be handled in the CSM framework \cite{Qin:2017lcd}.

\subsection{Evolution}
\label{sec:evolution}

Evolution equations play a markedly different role, whether we are considering the algebraic and numerical models of Sec. \ref{sec:DGLAPRegion} on the one hand, or the phenomenological one given in the previous section, on the other hand. Indeed, the latter is defined at a medium scale ($\mu_{\text{Ref.}}^2 = 1.9~ \textrm{GeV}^2$) and is already supplemented by strange quark and gluon distributions. On the contrary, the LFWFs-models are defined at a low scale (\emph{i.e} below $1\text{ GeV}^{2}$), where effective quarks are expected to be the relevant degrees of freedom to describe the pion. Strange quark and gluon distributions are then purely generated by evolution.

\begin{figure}[t]
	\begin{center}
          \includegraphics[scale=0.45]{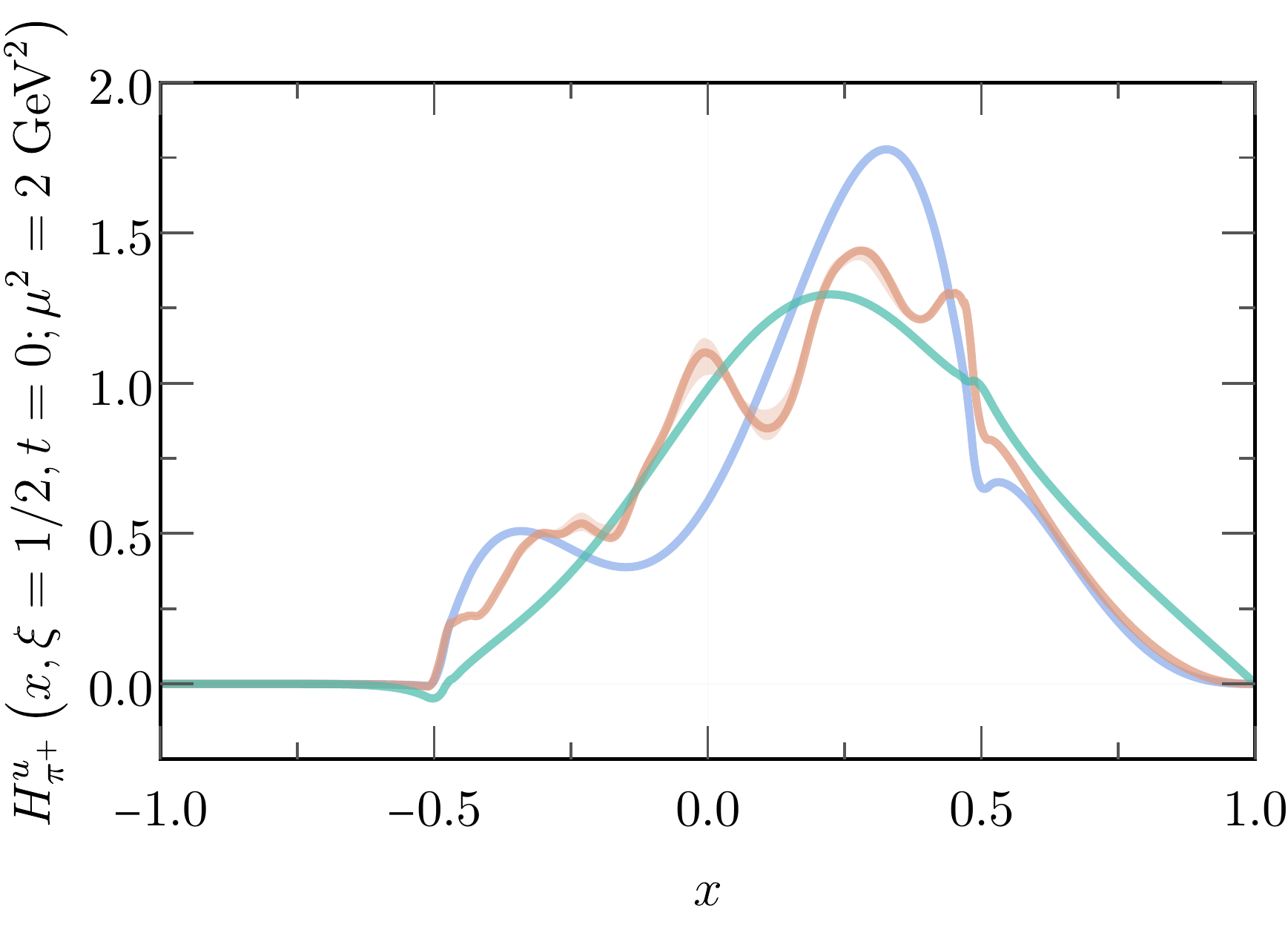} \\
          \includegraphics[scale=0.45]{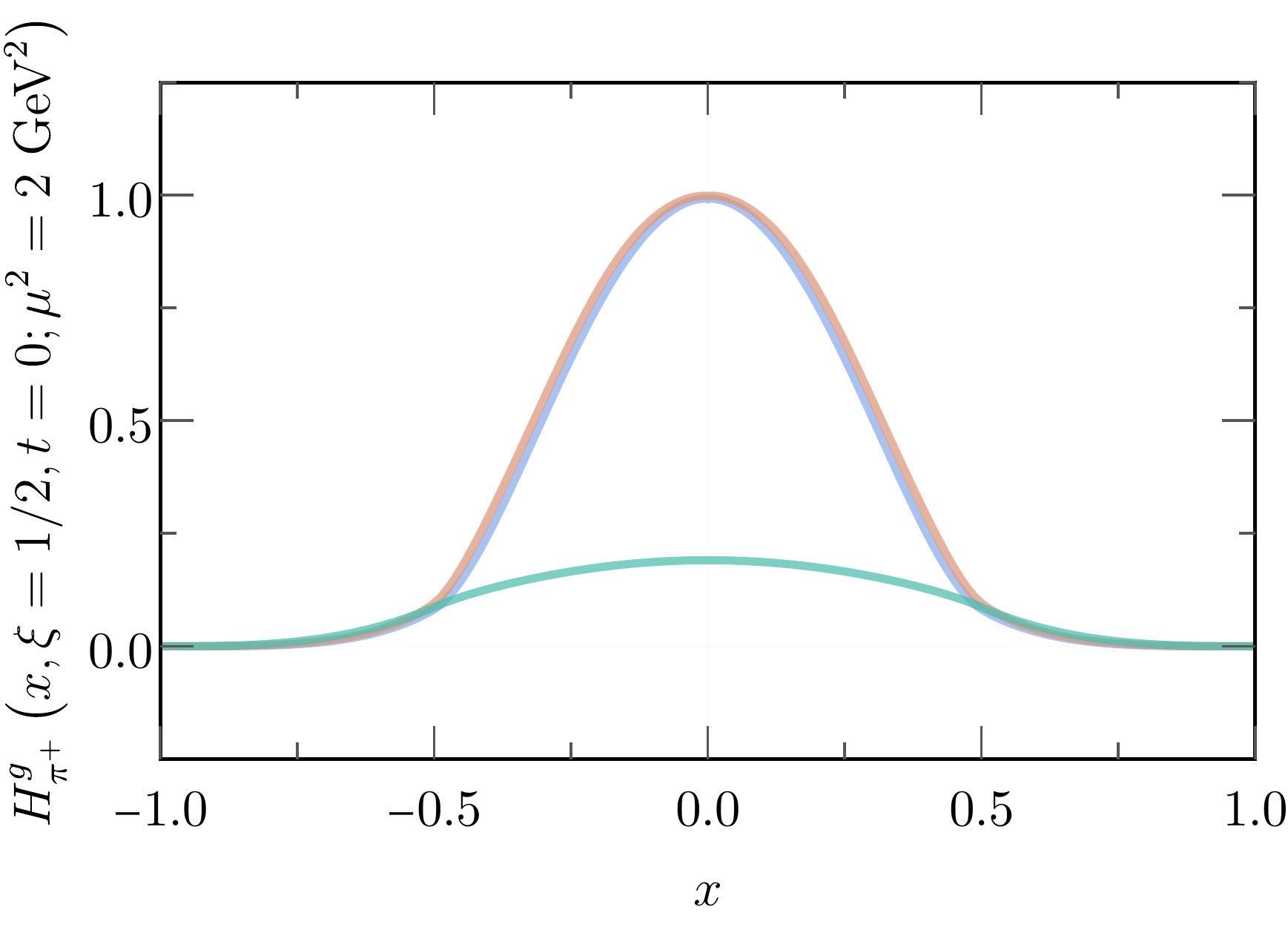}
        \end{center}
  \caption{\textit{Upper panel} - Comparison at $\mu^2 = 2~\textrm{GeV}^2$ of the three quark GPDs models for flavour-$u$ quarks at $t=0$ and $\xi = 1/2$. \textit{Lower panel} - Same thing for gluon GPDs.}
  \label{fig:EvolveGPDs}
\end{figure}

In order to be able to perform the evolution from a low enough scale, we follow the path highlighted in Refs. \cite{Rodriguez-Quintero:2018wma,Ding:2019lwe}. Namely, we use an effective coupling obtained from lattice-QCD and CSM analyses, which has the interesting property of not presenting a Landau pole and instead saturates in the infrared regime \cite{Binosi:2016nme,Rodriguez-Quintero:2018wma,Cui:2019dwv}. It has been shown that quark and gluon PDFs obtained through that procedure \cite{Ding:2019lwe,Cui:2020tdf} are consistent with pion's gluon PDFs computed on the lattice \cite{Fan:2021bcr,Chang:2021utv}. To do that, we employed the \texttt{PARTONS} software \cite{Berthou:2015oaw} in combination with the \texttt{Apfel++} evolution software \cite{Bertone:2013vaa,Bertone:2017gds,EvolutionPaper}. As an illustration of such a procedure, Fig. \ref{fig:EvolveGPDs} shows our three models at a scale of $\mu^{2}=2\text{ GeV}^{2}$. Notably, even if it stays null at $\xi =1$ in agreement with the soft-pion theorem, one can note that evolution provides a significant gluon contribution already for $\xi = 1/2$. In fact, at 2 GeV$^2$, the generated gluon distributions are much larger than the one obtained from the phenomenological xFitter/RDDA model. This can be explained by the small-$x$ behaviour of the respective PDFs. The xFitter collaboration has assumed that the gluon PDF behaves like $1/x$ at $\mu = 1.9\textrm{GeV}^2$, while the lattice and CSM gluon PDFs behaves at the same scale like $x^{-3/2}$.

Our CSM-based GPD models come with their uncertainty band, generated by the inversion of the Radon transform. Such uncertainty bands need to be propagated through the evolution. This is particularly relevant for the case of the numerical model. We assess this effect through the replica method: from the uncertainties estimated in Sec. \ref{subsec:Uncertainties}, we introduce Gaussian noise at the level of the corresponding DD and generate a set of 250 GPDs at the reference scale. Then, we employ \texttt{Apfel++} to evolve them up to $\mu^{2}=2\,\text{GeV}^{2}$. Thus we are able to generate a band estimating the uncertainties generated through the inverse Radon transform strategy and propagate it by evolution (Fig. \ref{fig:EvolveGPDs}). We note that the evolution procedure tends to reduce the size of the uncertainty band, stabilising the results at moderate and high scales.

\subsection{Compton Form Factors}
\label{sec:CFF}

\begin{figure}[b]
  \centering
  \begin{subfigure}{0.5\textwidth}
  \centering
  \includegraphics[width=0.9\textwidth]{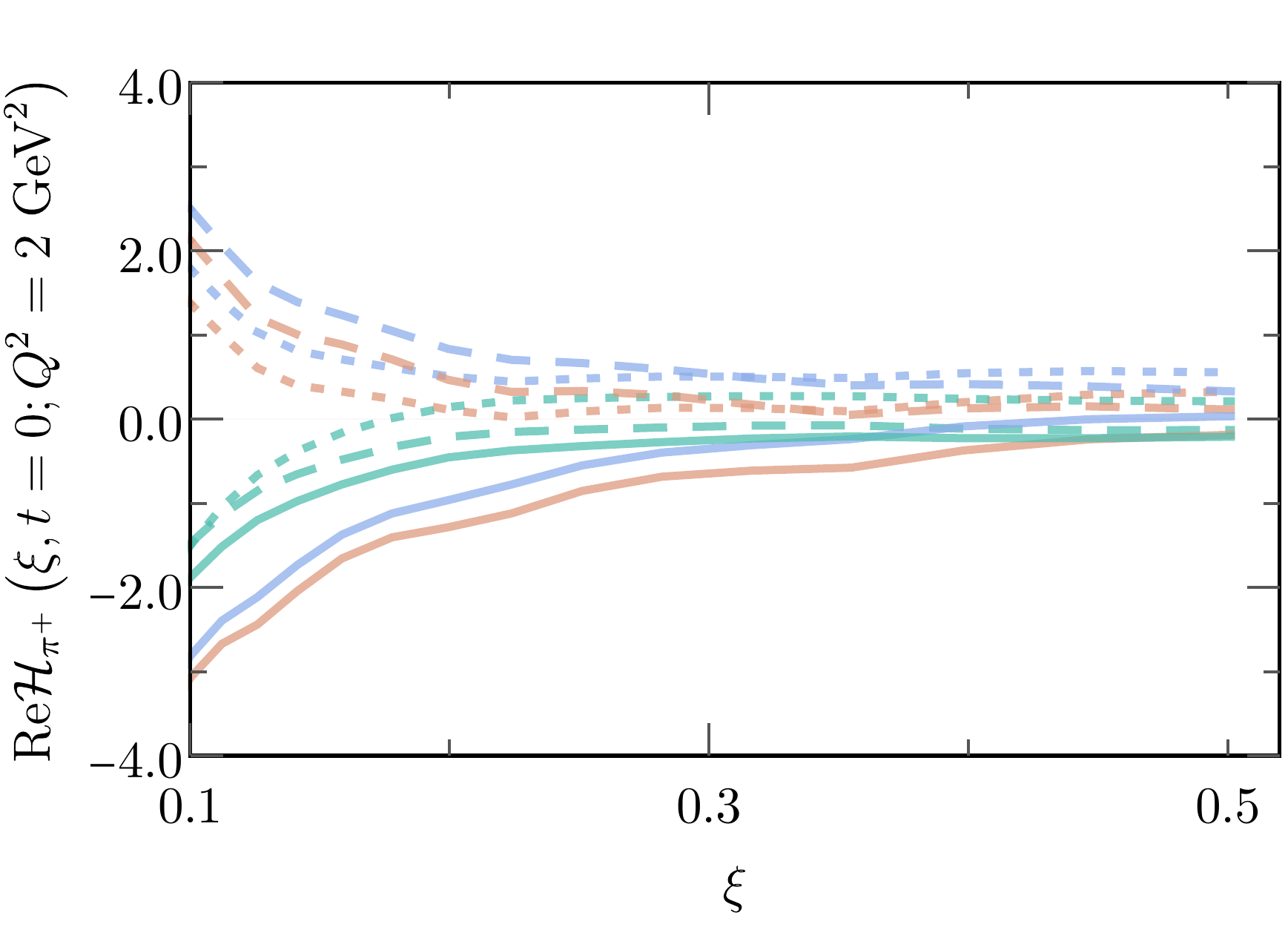}
  \end{subfigure}
  \begin{subfigure}{0.5\textwidth}
  \centering
  \includegraphics[width=0.9\textwidth]{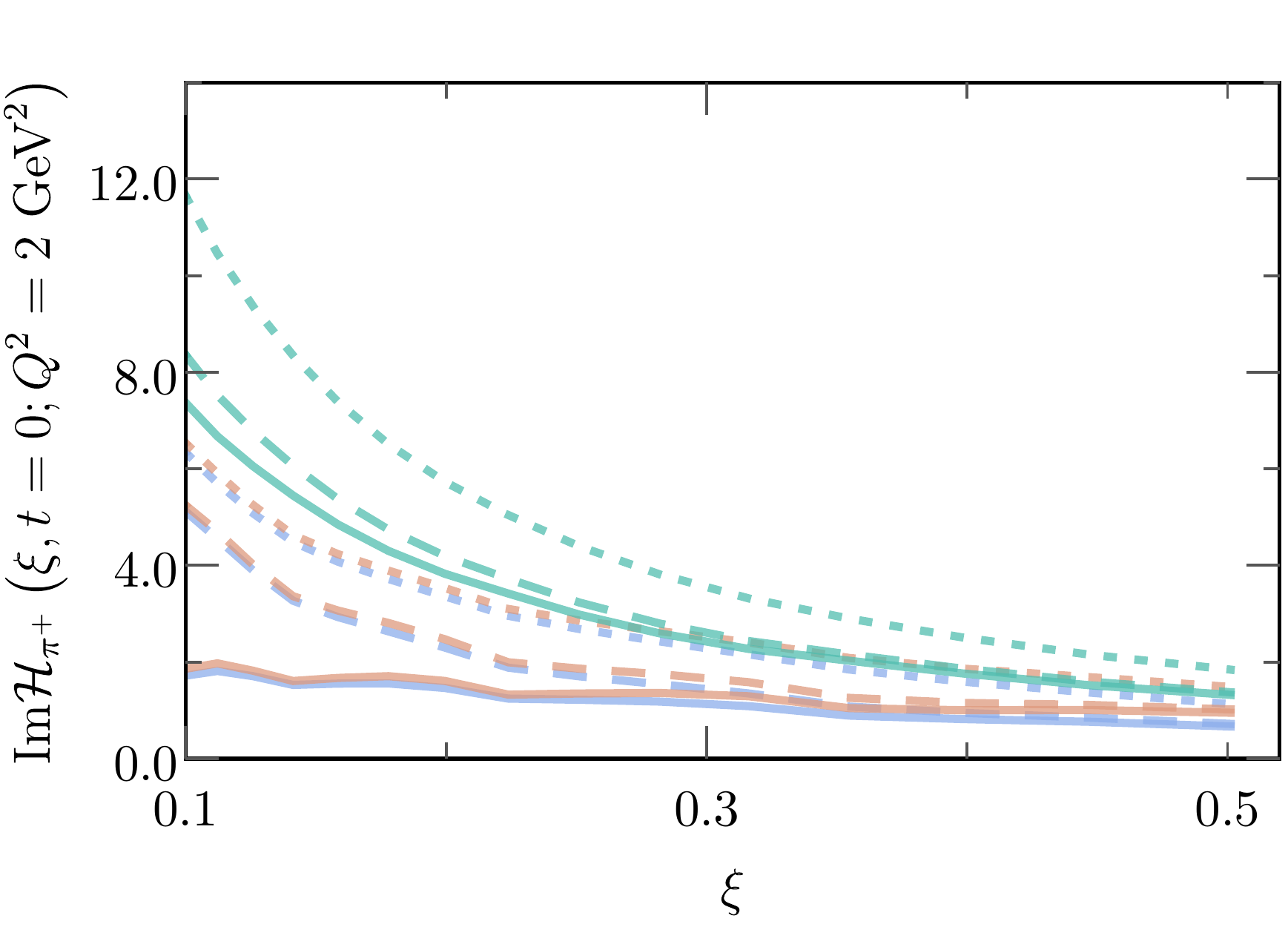}
  \end{subfigure}
  \caption{DVCS Compton Form Factors within the valence region, \textit{i.e.} $\xi\in\left[0.1,0.5\right]$. \textit{Upper panel} - Real part. \textit{Lower panel} - Imaginary part. \textit{Legend} - Blue line: algebraic model; brown line: numerical model; and green line: phenomenological model. Dotted line is the LO evaluation, dashed line the NLO without the gluon GPDs (see text), and solid line is the full NLO result.}
  \label{fig:CFFValence}
\end{figure}

Using our evolved GPD models, we computed the CFFs entering the description of DVCS using the formulae available in Refs. \cite{Pire:2011st,Moutarde:2013qs} and implemented in the \texttt{PARTONS} framework \cite{Berthou:2015oaw}. The results for the valence region are exhibited on Fig. \ref{fig:CFFValence}. Interestingly, the imaginary part of the CFF does not seem to be much sensitive to the differences between the algebraic and the numerical models, in all scenarii considered. The real part presents a sensitivity, but the latter remains small. More precisely, the richer physical content taken into account in the numerical model (including dynamical chiral symmetry breaking), merely generates a 10\% difference on the real part of the CFF, and almost nothing on the imaginary one.
Therefore, even in the valence region, DVCS seems to be poorly sensitive to the fine modelling assumptions, in agreement with \cite{Bertone:2021yyz}.
On the other hand, there is a clear difference between the phenomenological model on one side, and the Bethe-Salpeter derived ones on the other. Contrary to the latter, the former turns negative in the valence region at all considered orders of pQCD. This is an important outcome as such a region could be probed by the EicC project allowing one to discriminate between the two types of model presented here.

An additional comment that can be made on Fig. \ref{fig:CFFValence} is the key role of gluon GPDs, even in the valence region, thanks to the comparison done at NLO with and without taking them into account. For every model, gluons interfere with quarks, strongly reducing the imaginary part of the CFF when $Q^2$ remains low (a few $\textrm{GeV}^2$). The impact is more remarkable on the CSM-based model than on the phenomenological one, but the trend is the same. More dramatically, they also trigger a sign change in the real part of the amplitude for CSM-based models, and amplify the phenomenon in the case of the phenomenological model.

\begin{figure}[b]
  \centering
	\begin{subfigure}{0.5\textwidth}
	\centering  
	\includegraphics[width=0.9\textwidth]{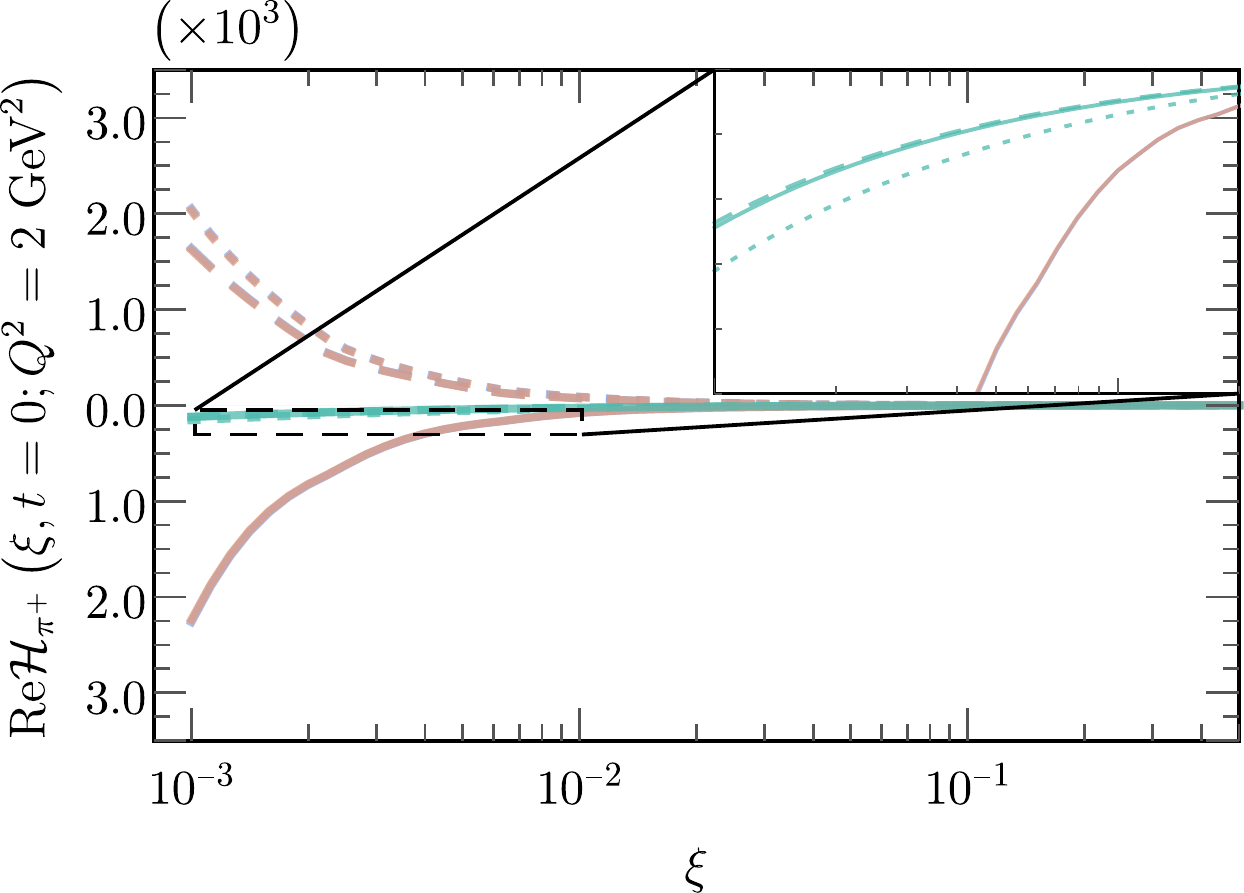}
	\end{subfigure}
	\begin{subfigure}{0.5\textwidth}
	\centering
	\includegraphics[width=0.9\textwidth]{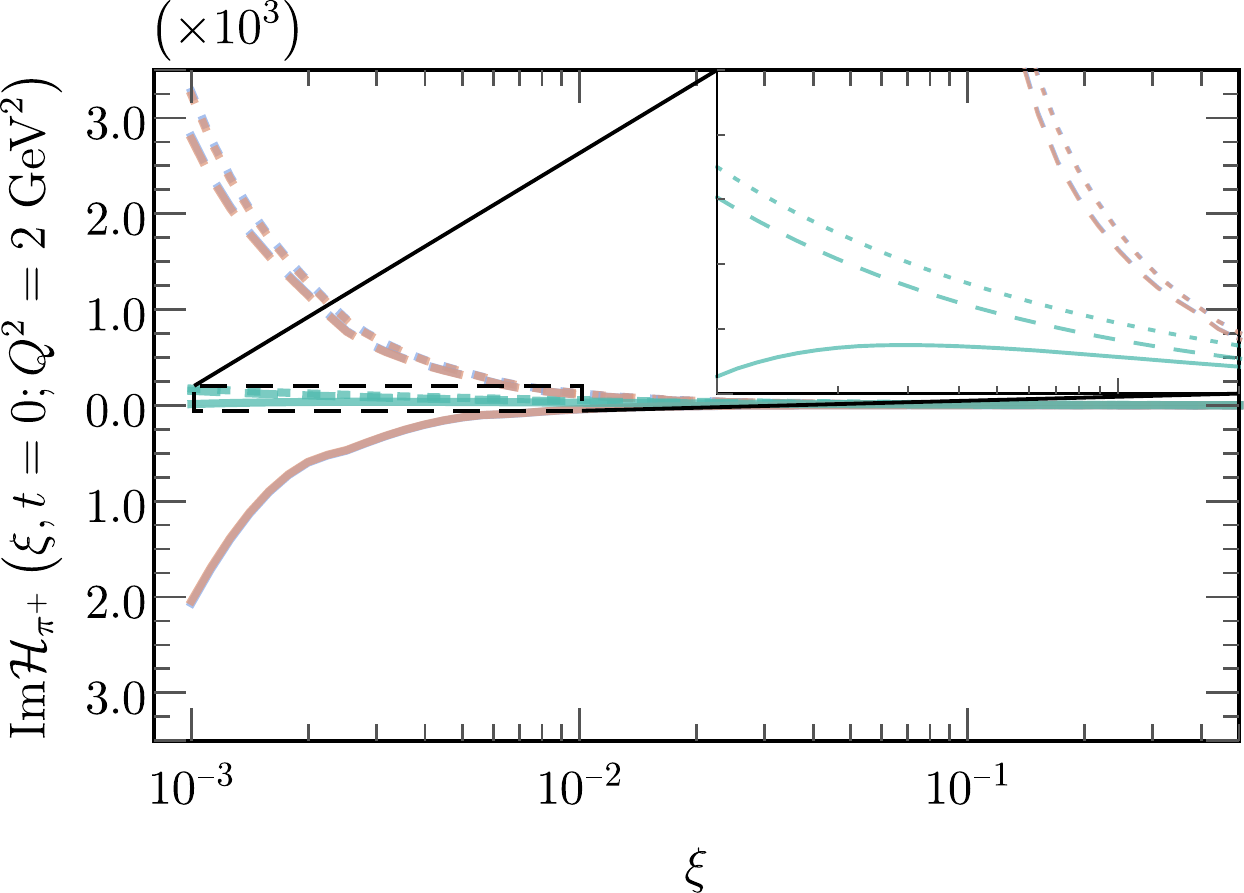}
	\end{subfigure}
  \caption{DVCS Compton Form Factors \textit{Upper panel} - Real part. \textit{Lower panel} - Imaginary part. \textit{Legend} - Brown line: numerical model; and green line: phenomenological model. Dotted line is the LO evaluation, dashed line the NLO without the gluon GPDs (see text), and solid line is the full NLO result. The algebraic model yields essentially indistinguishable results to the one of the numerical model.}
  \label{fig:CFFEIC}
\end{figure}

Outside of the valence region, the picture is modified as displayed in Fig. \ref{fig:CFFEIC}. This time, CSM-based models clearly yield a much larger CFF than the phenomenological one. This can be explained by the differences in the small-$x_B$ behaviour of the two type of models, as mentioned previously. In the low $\xi$ region (or equivalently low $x_B$ region), the dominance of the gluon is even more obvious as this time even the sign of the imaginary part of the RDDA-model's CFF is changed. At such low $\xi$ the differences between algebraic and numerical CSM-models become irrelevant, as evolution together with the convolution of the perturbative kernel wash out the differences. This is again in agreement with the finding of Ref. \cite{Bertone:2021yyz} on the DVCS deconvolution problem. 

We note that gluon dominance at low $\xi$ yields a CFF roughly behaving as $1/\xi^b$ with $b\approx 1.4$ in the case of the numerical model. Such a behaviour remains compatible with DVCS dispersion relations with one subtracted constant \cite{Diehl:2007jb}. However, we highlight that our study is a pure NLO one with no small-$\xi$ resummation being taken into account. The latter may have an important impact on the $\xi$ behaviour of the CFF in the low-$\xi$ region. We left this point for a future work. 

\begin{figure}[t]
  \centering
  \begin{subfigure}{0.5\textwidth}
  \centering
  \includegraphics[width=0.9\textwidth]{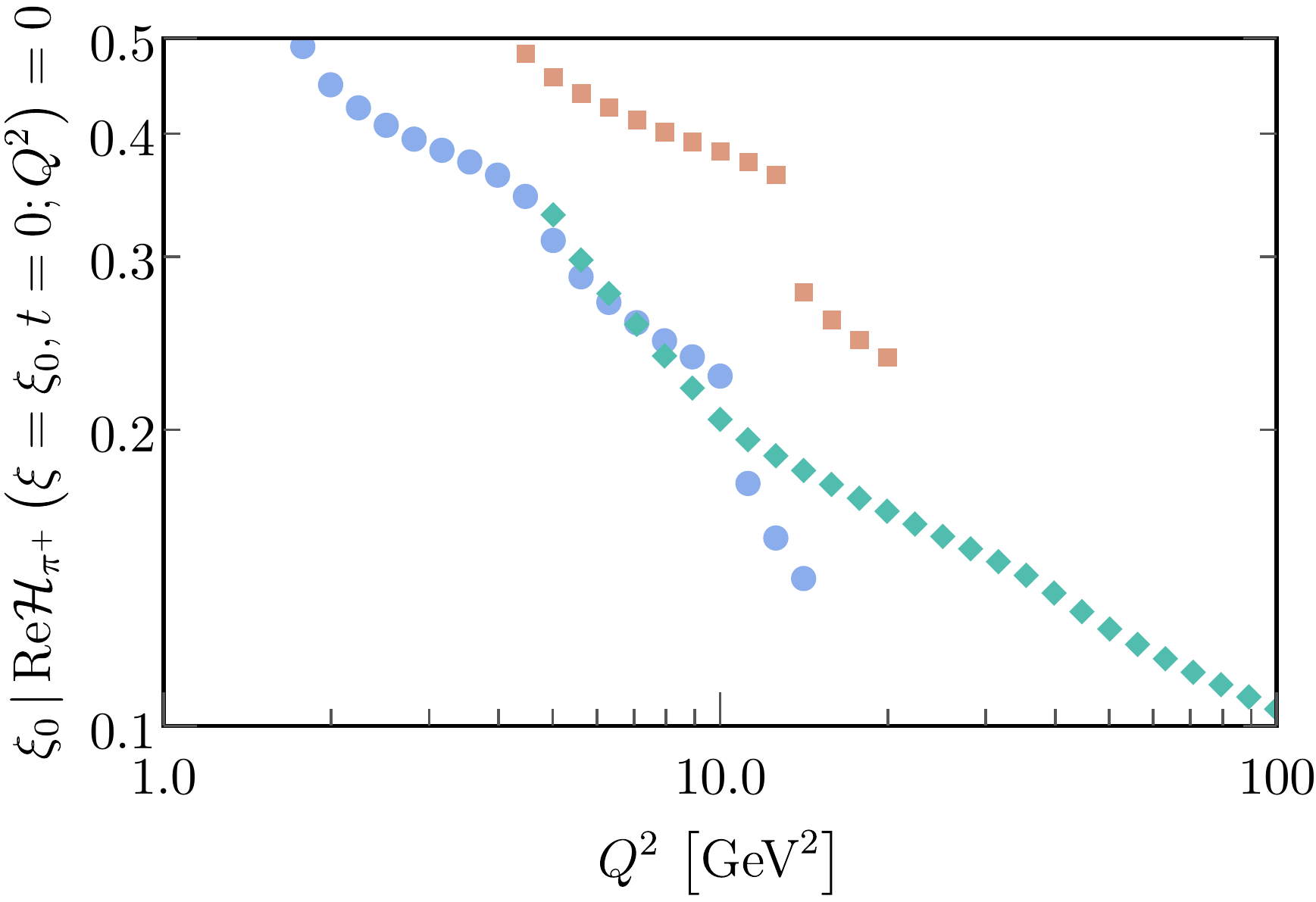}
  \end{subfigure}
  \begin{subfigure}{0.5\textwidth}
  \centering
  \includegraphics[width=0.93\textwidth]{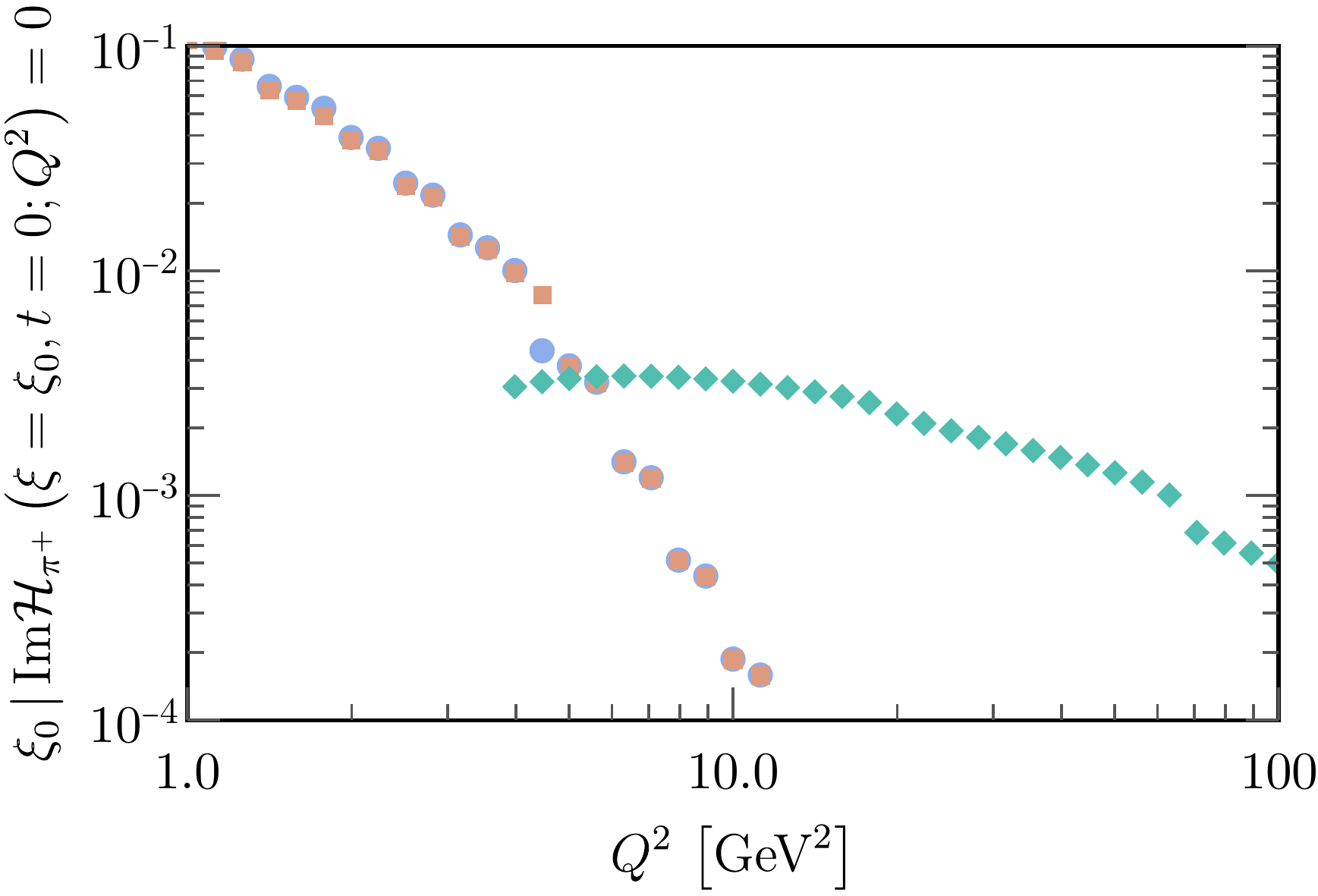}
  \end{subfigure}
  \caption{\textit{Upper panel} - $Q^{2}$ displacement of $\xi_{0}|\textrm{Re}\mathcal{H}_{\pi^+}(\xi_0,t=0; Q^2)=0$. \textit{Lower panel} - $Q^{2}$ displacement of $\xi_{0}|\textrm{Im}\mathcal{H}_{\pi^+}(\xi_0,t=0; Q^2)=0$. \textit{Legend} - Blue dots: algebraic model; brown squares: numerical model; and green rhombus: phenomenological model. As $Q^2$ increases, $\xi_0$ decreases, pushing the gluon dominance toward lower values of $x_B$, as we would naively expect from a perturbative expansion.}
  \label{fig:CFFZeroCrossing}
\end{figure}

Finally, we would like to stress an interesting feature of the present study; namely that due to gluon GPDs, the real and imaginary parts of the CFF change sign at given values $\xi_0$ which run with $Q^2$. This is illustrated on Fig. \ref{fig:CFFEIC} where we see that NLO corrections without taking gluon GPDs into account keep the same sign as the LO results. Only the inclusion of gluon GPDs triggers a zero crossing in the real and imaginary part of the CFF. The $Q^2$ dependence can be understood in the following way: if at some scale $Q^2_1$ and at some skewness $\xi_{1}$ the imaginary part of the CFF is dominated by gluons (\emph{i.e.} it is negative), increasing the scale should reduce the impact of NLO corrections. Therefore, one expects that at some sufficiently high scale $Q^2_2$ the CFF evaluated at $\xi_1$ turns positive. In other words, the zero crossing is shifted toward lower values of $\xi$ when $Q^2$ increases. This is precisely what we observe, as displayed on Fig. \ref{fig:CFFZeroCrossing}. Interestingly, our phenomenological model undergoes such zero crossing over the entire $Q^{2}$ range studied here. On the other hand, the two CSM-based models exhibit an abrupt step at $Q^{2}\simeq 10-20\text{ GeV}^{2}$, for both real and imaginary parts of the CFF.
We also note that, contrary to other features of CFFs, the zero-crossing in the real part of the latter allows one to clearly differentiate between the algebraic and numerical models. This might be an experimental signal able to distinguish between the different physical assumptions used.
However, the real part of the CFF is usually more difficult to extract experimentally and thus the study of the consequences of this zero crossing on DVCS cross section and asymmetries is left for a future work. 

\section{Conclusion}
\label{sec:Conclusion}

As we illustrated, the path to build GPD models able to fulfil by construction all the theoretical constraints which apply to these matrix elements, is tough. Nevertheless, we showed a way to go, taking advantage of ab-initio computations of the Bethe-Salpeter wave function of the pion. Combining the mathematical structure of the overlap of LFWFs together with the properties of the inverse Radon transform, we built a pion quark GPD able to fulfil by construction all the required theoretical properties. We also take the opportunity to improve the numerical solution presented previously in \cite{Chouika:2017dhe,Chouika:2017rzs}, allowing us to assess the numerical uncertainties triggered by the ill-conditioned character of the inverse Radon transform, and highlight the filtering character of the evolution kernel. This allowed us to build for the first time a consistent GPD from a numerical solution of the Bethe-Salpeter equation, something which was long sought (see for instance Refs. \cite{Mezrag:2014tva,Mezrag:2014jka,Mezrag:2016hnp,Mezrag:2015mka,Shi:2018mcb})

Our work also validates \texttt{PARTONS} \cite{Berthou:2015oaw} as a modular tool able to bridge the gap between non-perturbative QCD practitioners, either using continuum or Lattice techniques, and physical observables related to the 3D structure of hadrons. We stress once again the crucial role played by evolution in this study that we exploited through the \texttt{Apfel++} library \cite{Bertone:2013vaa,Bertone:2017gds}. Its combined usage with \texttt{PARTONS} made possible to compute CFFs on a large range of $\xi$, $t$ and $Q^2$ kinematical points covering both the kinematical regions explored at EIC and EicC. This has revealed two interesting features: i) the remarkable sign change in the imaginary part of the CFF, testifying of the importance of the gluon GPD and ii) a feasibility study of accessing experimentally pion GPDs at EIC and EicC using the Sullivan process \cite{Sullivan:1971kd} (see also \cite{Huber:2008id,Qin:2017lcd,Aguilar:2019teb}), something advocated in the EIC Yellow Report \cite{AbdulKhalek:2021gbh}. This study is presented in a dedicated paper.  

Finally, we mention that our study on the pion has consequences at the level of the nucleon. Indeed, the validation of the computing chain for the pion holds for the nucleon, highlighting the fact that \texttt{PARTONS} is ready for phenomenological studies of DVCS at NLO. On top of this, the physical behaviour of the CFF, especially the sign change of the imaginary part, might well be something that will be observed also for the nucleon. Further study in that direction must be performed before the EIC can start accumulating data.

\appendix
\section{Invertibility of $\mathcal{R}^{T}\mathcal{R}$}
\label{app:Invertibility}

In Sec. \ref{sec:Extension}, we addressed the problem of computing DDs by solving a squared linear system whose matrix is written as:
\begin{equation}
(\mathcal{R}^{T}\mathcal{R})_{jk} = \sum_i R_{ij} R_{ik}
\end{equation}
where $\mathcal{R}_{ij}$ is the contribution of the element $j$ to the integral over the line $x_{i}-\beta-\alpha\xi_{i}=0$ with $\left(x_{i},\xi_{i}\right)$ in the DGLAP region. Its inversion is a necessary step for the solution of the inverse Radon transform problem. Thus the condition of maximal rank for the matrix $\mathcal{R}^{T}\mathcal{R}$ must be fulfilled.

As we shall prove through this appendix, such a condition is unavoidably met if the Radon transform matrix, $\mathcal{R}$, has maximal rank; a condition which, as discussed in Sec. \ref{subsec:Discretization}, can be assumed to be true without loss of generality.

Prior to our proof for the invertibility of $\mathcal{R}^{T}\mathcal{R}$, we must present the two central pieces of our arguments:
\\

\noindent \textbf{1.} \textit{Rank-nullity theorem}

Let $V,W$ be finite dimensional $\mathbb{F}$-vector spaces and $T:V\rightarrow W$ a linear application. Then, the \textit{rank-nullity theorem} states:
\begin{equation}
\begin{array}{rcl}
\displaystyle dim_{\mathbb{F}}V & \displaystyle = & \displaystyle \dim_{\textbf{F}}T\left(V\right)+dim_{\textbf{F}}ker\left(T\right) \\
& & \\
& = & Rank\left(T\right)+dim\mathcal{N}\left(T\right)\\ \end{array}
\end{equation}
with $\mathcal{N}\left(T\right)$ denoting the null-space of the application.

In particular, for a matrix $\mathcal{A}\in\mathcal{M}_{m,n}\left(k\right)$, with $m\geq n$:
\begin{equation}
n=Rank\left(\mathcal{A}\right)+dim\mathcal{N}\left(\mathcal{A}\right)
\end{equation}
from which one can straightforwardly deduce that,
\begin{equation}\label{eq:NulRank}
dim\mathcal{N}\left(A\right)=0\Leftrightarrow Rank\left(\mathcal{A}\right)=n
\end{equation}
\textit{i.e.}, the matrix $\mathcal{A}$ has maximal rank.

Therefore, in the particular situation where the matrix $\mathcal{A}_{n}$ is squared, \emph{i.e.} $m=n$, the condition $dim\mathcal{N}\left(\mathcal{A}_{n}\right)=0$ implies that such matrix has maximal rank and thus, by means of \textit{Rouché-Frobenius theorem}, that such matrix is invertible:
\begin{equation}\label{eq:NulInv}
dim\mathcal{N}\left(\mathcal{A}_{n}\right)=0\Leftrightarrow\exists\thinspace\mathcal{A}_{n}^{-1}\in\mathcal{M}_{n}|\mathcal{A}_{m}\mathcal{A}_{n}^{-1}=\mathcal{A}_{n}^{-1}\mathcal{A}=\mathbb{I}_{n}
\end{equation}

\noindent \textbf{2.} $\mathcal{N}\left(A\right)=\mathcal{N}\left(\mathcal{A}^{T}\mathcal{A}\right)$

Once again, let us consider an arbitrary matrix $\mathcal{A}\in\mathcal{M}_{m,n}\left(k\right)$, with $m\geq n$ and a vector $x\in\mathcal{N}\left(A\right)$. Applying $\mathcal{A}^{T}\mathcal{A}\in\mathcal{M}_{n}\left(k\right)$ on it:
\begin{equation}
\mathcal{A}^{T}\mathcal{A}x=\mathcal{A}^{T}\textbf{0}=\textbf{0}
\end{equation}
where the first identity follows from the definition of $\mathcal{N}\left(\mathcal{A}\right)$, it immediately implies that $x\in\mathcal{N}\left(\mathcal{A}^{T}\mathcal{A}\right)\Rightarrow \mathcal{N}\left(A\right)\subset\mathcal{N}\left(\mathcal{A}^{T}\mathcal{A}\right)$.

Equivalently consider a vector $x\in\mathcal{N}\left(\mathcal{A}^{T}\mathcal{A}\right)$. Then,
\begin{equation}
\left(\mathcal{A}^{T}\mathcal{A}\right)x=\textbf{0}\Rightarrow x^{T}\left(\mathcal{A}^{T}\mathcal{A}\right)x=x^{T}\textbf{0}=\textbf{0}
\end{equation}
and thus,
\begin{equation}
x^{T}\left(\mathcal{A}^{T}\mathcal{A}\right)x=\left(\mathcal{A}x\right)^{T}\left(\mathcal{A}x\right)=\left|\left|\mathcal{A}x\right|\right|^{2}=0
\end{equation}
where $\left|\left|\cdot\right|\right|$ denotes the vector norm.

Because $\mathcal{A}$ is different from the null-operator, it follows that:
\begin{equation}
\left|\left|\mathcal{A}x\right|\right|^{2}=0\Rightarrow \mathcal{A}x=\textbf{0}
\end{equation}
thus $x\in\mathcal{N}\left(A\right)\Rightarrow\mathcal{N}\left(\mathcal{A}^{T}\mathcal{A}\right)\subset\mathcal{N}\left(\mathcal{A}\right)$.

The combination of these two results imply
\begin{equation}\label{eq:NullEquiv}
\mathcal{N}\left(A\right)=\mathcal{N}\left(\mathcal{A}^{T}\mathcal{A}\right)
\end{equation}

Keeping this in mind let us turn to the specific problem we are involved with. Consider $\mathcal{R}\in\mathcal{M}_{m\times n}\left(\mathbb{R}\right)$, the Radon transform matrix of Sec. \ref{sec:Extension}, with $m\geq n$. And the matrix $\mathcal{R}^{T}\mathcal{R}\in\mathcal{M}_{n}\left(\mathbb{R}\right)$, where $\mathcal{R}^{T}$ stands for the transposed Radon transform matrix.

By hypothesis, $Rank\mathcal{R}=n$, as discussed through Sec. \ref{subsec:Discretization}. Therefore, by means of Eq. \eqref{eq:NulRank}, $dim\mathcal{N}\left(\mathcal{R}\right)=0$. Furthermore, relation \eqref{eq:NullEquiv} guarantees that $dim\mathcal{N}\left(\mathcal{R}^{T}\mathcal{R}\right)=dim\mathcal{N}\left(\mathcal{R}\right)=0$. Then, through \eqref{eq:NulInv}, the matrix $\left(\mathcal{R}^{T}\mathcal{R}\right)^{-1}$ exists.

\begin{acknowledgments}
We are grateful to Qin-Tao Song for kindly providing us the GFF data obtained from pion GDA. We would like to thank P. Barry, H. Dutrieux, T. Meisgny, B. Pire, K. Raya, C.D. Roberts, P. Sznajder and J. Wagner for interesting discussions and stimulating comments. This work is supported by University of Huelva under grant EPIT-2019 (J.M.M.C). F.S., J.R.Q. and J.S. acknowledge support from Ministerio de Ciencia e Innovación (Spain) under grant PID2019-107844GB-C22; Junta de Andalucía, under contract No. operativo FEDER Andalucía 2014-2020 UHU-1264517 and P18-FR-5057 and PAIDI FQM-370. This project was supported by the European Union's Horizon 2020 research and innovation programme under grant agreement No 824093. This work is supported in part in the framework of the GLUODYNAMICS project funded by the "P2IO LabEx (ANR-10-LABX-0038)" in the framework "Investissements d’Avenir" (ANR-11-IDEX-0003-01) managed by the Agence Nationale de la Recherche (ANR), France.
\end{acknowledgments}

\bibliography{Bibliography}
\bibliographystyle{unsrt}

\end{document}